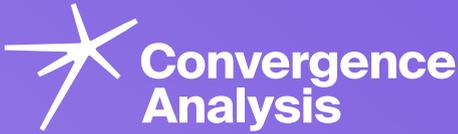

# AI Model Registries

## A Foundational Tool for AI Governance

ELLIOT MCKERNON, GWYN GLASSER, DERIC CHENG, GILLIAN HADFIELD

SEPTEMBER 2024

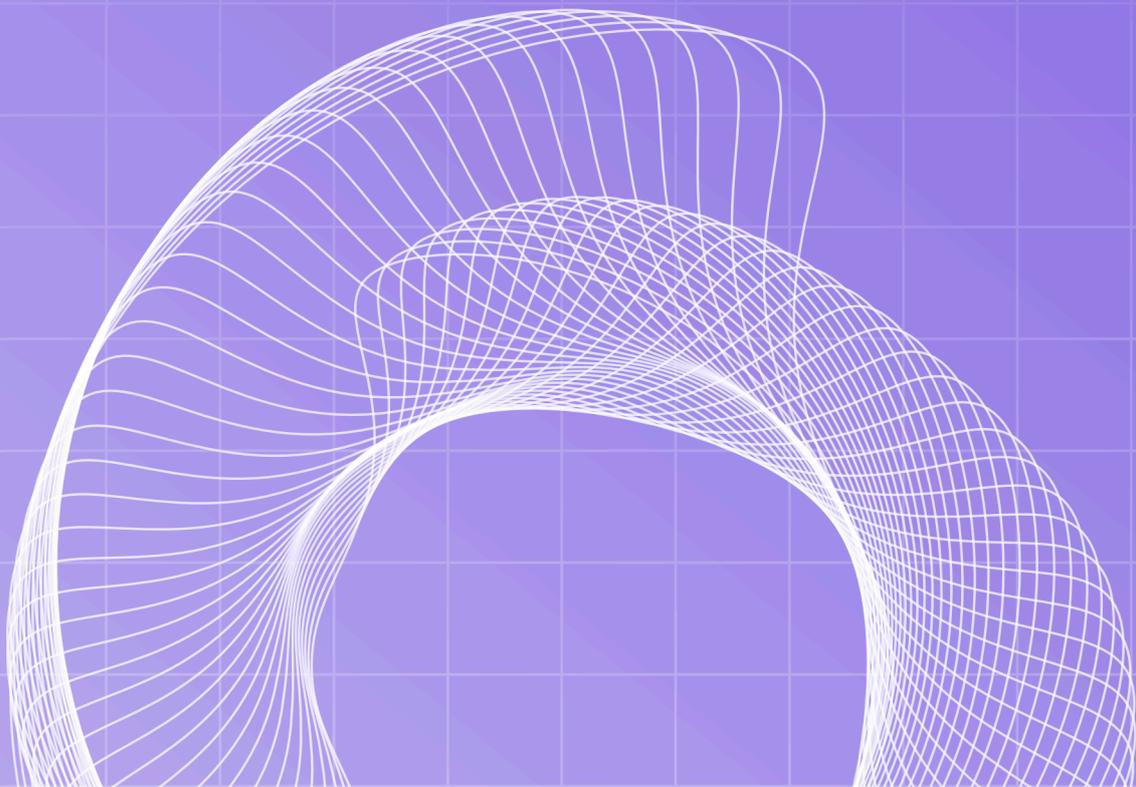

# Table of Contents

# Abstract

In this report, we propose the implementation of national registries for frontier AI models as a foundational tool for AI governance. We explore the rationale, design, and implementation of such registries, drawing on comparisons with registries in analogous industries to make recommendations for a registry that's efficient, unintrusive, and which will bring AI governance closer to parity with the governmental insight into other high-impact industries. We explore key information that should be collected, including model architecture, model size, compute and data used during training, and we survey the viability and utility of evaluations developed specifically for AI. Our proposal is designed to provide governmental insight and enhance AI safety while fostering innovation and minimizing the regulatory burden on developers. By providing a framework that respects intellectual property concerns and safeguards sensitive information, this registry approach supports responsible AI development without impeding progress. We propose that timely and accurate registration should be encouraged primarily through injunctive action, by requiring third parties to use only registered models, and secondarily through direct financial penalties for non-compliance. By providing a comprehensive framework for AI model registries, we aim to support policymakers in developing foundational governance structures to monitor and mitigate risks associated with advanced AI systems.

### Collaborators and Acknowledgements

This paper was written in collaboration with Mariano-Florentino (Tino) Cuéllar and Tim O'Reilly.

We'd like to thank the following people for their insightful feedback and input on this project: Scott Singer, Justin Bullock, Zershaaneh Qureshi, Tommy Shaffer Shane, and James Gealy.

# Executive Summary of Our Proposal

**In this report, we propose that national governments should implement AI model registries as a foundational tool for AI governance.** By *model registry*, we mean a centralized *database of frontier AI models* that includes standard commercial and specific safety-relevant information about these models and their deployers. Developers would be required to report any qualifying models and their information to the registry before public deployment.

In Parts I, II, and III, we explore and make recommendations on the purpose of such a registry, what information it should store, and how to practically implement and administer it, respectively.

**In this executive summary of our proposal we provide a concise, high-level summary of each of our conclusions, without argumentation, analysis, or evidence.** To understand why we make each of these specific recommendations, we encourage readers to read the full section on each topic.

## The Case for a Model Registry

AI model registries can serve as a foundational lever to increase regulatory visibility, support legal action, and manage societal risks. In other industries, registries successfully serve this same purpose for products and services associated with notable economic impacts or risks to society, as we detail in *Registries are a basic, common governance tool*. However, while some nations are taking early steps to develop model registries, as we detail in *What AI model registries currently exist?*, the current standards for frontier AI registration are not yet substantial enough to bring AI oversight into parity with other industries.

We identify four high-level objectives that motivate the adoption of frontier model registries:

- A registry will **facilitate the monitoring of frontier AI technology**, providing governments with increased regulatory visibility into the capabilities and risks of leading AI models.

- A registry will provide a **key mechanism for regulatory enforcement** of AI models, enabling governments to accurately pinpoint models subject to regulation.

- A registry will **enable the development of new regulation** and serve as a foundational governance hub, allowing governments to classify models and create regulation based on specific capabilities or characteristics.

- A registry will **foster public sector field-building** by promoting the use of

common standards, providing structured information on AI for policymakers, and encouraging the development of the technical skills and knowledge required to manage AI systems.

Crucially, a registry can achieve these four important goals efficiently and without hobbling innovation. We elaborate on these benefits in *What value does a model registry provide to governments?*.

## Proposed Design of a Model Registry

Based on our research detailed in Part II, we propose that an effective AI model registry should adhere to the following design principles to achieve the goals listed above:

- A model registry should be **minimal**, and aim to *only* require the information needed to fulfill the described purposes.
- A model registry **should not include licensing requirements or mandatory standards.** It should primarily consist of reporting existing information about an AI model, and require minimal additional overhead for developers.
- A model registry should be **interoperable** and conform to international standards that minimize the regulatory burden on registry administrators and AI developers.
- The bar for inclusion into a model registry should be **low enough to capture the next generation of highly capable frontier models, but above the current generation of models** (those deployed before the publication of this report).
- Models should be required to be **registered prior to deployment**.
- The registry should support **categorizing models into families**, and allow developers to maintain the model information for only the most capable models in each key measurable dimension to minimize overhead.
- Developers should be required to revisit their registry entries twice a year, either confirming that the information remains accurate or updating it to reflect any changes.
- An effective model registry should contain information including:
    - Basic information on the developing organization
    - Open-source status of the model
    - Model size in parameters
    - Compute used during training, retraining, and post-training
    - Training data: amount, type, and provenance
    - A high-level description of model architecture
    - General information about the hardware used for development
    - A description of the security standards protecting key components of the AI model

- The mechanism and results of any model evaluations or benchmarks conducted by the developer
- A description of the functions of the model
- A summary of post-deployment monitoring techniques used.

## Proposed Implementation of a Model Registry

Based on our research detailed in Part III, we propose that an effective AI model registry should meet the following implementation principles:

- A model registry should be **enforced by implementing a system to fine AI developers** a percentage of annual turnover for non-compliance.
- A model registry should **require third-party users of frontier AI models to verify that those models have been registered.**
- A model registry should be **overseen directly by governments with minimal outsourcing** to third-parties.
- A model registry should be **implemented at the national level**, but remain interoperable with international standards.
- A model registry should be **pragmatically confidential and secure.**

## Structure of the report

In Part I, we explore why AI models require greater governance and introduce model registries as a potential governance tool. We explore the benefits a registry could provide to governments and society and the risks that should be mitigated in designing and implementing a model registry.

In Part II, we research and make recommendations on how to design an effective registry: which models should qualify for inclusion on the registry, and what information developers should submit to the registry about their models.

In Part III, we research and make recommendations on how to practically implement an effective registry: how it should be administered, whether its information should be public or private, and how to ensure developers share accurate information.

For each topic, we share our research, weigh benefits and risks, and conclude by making specific recommendations.

# Part I - Motivation for a Model Registry

In this section, we argue that AI model registries are a critical foundational tool for AI governance, providing governments with regulatory visibility, enabling enforcement, and supporting the development of future regulation. We conclude that while some early registry efforts exist, current implementations are insufficient to achieve the full range of benefits a comprehensive registry could provide, such as facilitating monitoring of frontier AI technology, providing accountability mechanisms, and fostering public sector expertise in AI governance.

## What value does a model registry provide to governments?

AI has advanced dramatically in the last decade, leading to the proliferation of generative AI and large language models like ChatGPT that are capable of producing images, audio, and text at qualities near or surpassing many humans. The resources invested in AI have been growing exponentially for decades with global corporate investment peaking at $337 billion in 2021[1].

Like all emerging technologies, frontier AI brings both significant opportunities and risks. The recent US Executive Order on AI[2] and the EU AI Act[3] both highlight this dual promise of AI, and though experts disagree on the severity of advantages and harms AI will bring, there is an consensus that AI will have a huge impact on our economies and societies in the coming years[4,5].

However, unlike other technologies, AI development is largely unregulated and opaque to policymakers and the public. Governments lack insight into the capabilities and risks of models, and into how these models are developed and deployed, depriving them of the capacity to predict and mitigate safety issues[6]; currently, governments must rely on what information AI developers volunteer to share. No other industry has such a large impact or role within the economy without major oversight to ensure safety, and the impact of AI will only magnify as development accelerates[7].

As laid out in our recent report on the State of the AI Regulatory Landscape[8], the EU, US, and PRC are beginning to monitor and regulate AI development, but these efforts are limited in scope and still developing. Governments will need to take further action to make the most of the opportunities offered by AI and minimize risks. To do that, they will need to understand AI developers, their models, and models' capabilities. They will require greater insight into AI development.

**To give governments basic insight into AI development on par with similar industries, and to provide them with the information necessary to design high-quality, evidence-based AI policy, we propose governments adopt a national AI model registry.**

## Registries are a basic, common governance tool

In other industries, registration is often the first step in enabling legal action against a responsible entity. A basic example is corporate registration, which creates accountability by making a corporation visible and subject to suit[9]. The registration process requires identification of a person or entity as an agent for the company. This agent is authorized to accept service of process, which is how a legal action, including public enforcement, is initiated. Registries in high-risk industries often have higher reporting requirements to increase regulatory visibility into the respective domain and allow for a range of government interventions to mitigate risk[10].

These are typical requirements in many industries in many countries, and yet frontier AI models and developers face minimal if any reporting requirements, as we discuss in _What AI model registries currently exist?_. To emphasize this point, we'll frequently compare our recommendations to reporting requirements in other industries such as food, drugs, weapons, chemical manufacturing, and more. In this section, we'll go through some of these examples in more detail.

### The FDA

The US Food and Drug Administration is one of country's oldest consumer protection agencies with origins in the latter half of the 19th century[11]. Today, the FDA maintains extensive registries for food products and producers, drugs, clinical trials, medical equipment, and more.

- Food facilities handling products for US consumption must register, providing their name, address, food categories, and contact details[12]. Drug manufacturers, repackers, and relabelers register their establishments, listing facility information, drug products, and manufacturing activities.

- Medical device companies register with similar establishment details, but also include specific device listings and performed activities. For safety monitoring, the FDA employs FAERS for drugs and biologics, MAUDE for medical devices, and VAERS for veterinary products[13]. These systems collect detailed reports of adverse events, including product information, event description, patient outcomes, and reporter details from healthcare professionals, consumers, and manufacturers.

- The UDI database for medical devices requires more technical information, including unique device identifiers, product names, models, and versions. This system aims to precisely track devices throughout their lifecycle.

Each registry is tailored to its specific industry, with reporting requirements designed to provide the FDA with comprehensive oversight while balancing the need for efficiency in data collection and management. These also support pre-market approval, pre- and post-market monitoring, and incentives to encourage innovation (such as temporary exclusive rights to manufacture newly developed drugs[14]).

### REACH

The EU REACH regulation requires chemical suppliers to register the substances they manufacture in the EU to the European Chemicals Agency[15]. It covers all substances manufactured or imported in the EU above one tonne per year. Compared to FDA registries, REACH is broader in scope and demands more extensive safety data and risk management across the entire chemical supply chain. Registrants must provide detailed information, including chemical identity, use details, safety data, and toxicological information. The European Chemicals Agency and Member States evaluate submissions, with special authorization required for substances of high concern. REACH also allows for EU-wide restrictions on chemicals posing significant risks.

### Nuclear material

The US Nuclear Regulatory Commission (NRC) oversees a national registry of products containing nuclear material that includes "information on the sources and devices, such as how they are permitted to be distributed and possessed (specific license, general license, or exempt), design and function, radiation safety, and limitations on use[16]." The NRC also provides platforms for incident reporting, conducts regular inspections of manufacturers, and has well-developed enforcement mechanisms for ensuring compliance to safety standards.

### Federal Select Agent program

The US Federal Select Agent program identifies a specific list of regulated substances, and maintains a national database of organizations handling these substances. Registered entities must disclose general information about the industry they operate in, contact information for a responsible entity, which select agents will be handled, who has access to them, and how a range of safety and security measures are implemented[17]. Disclosed information is confirmed with inspections, and background checks are conducted on key individuals in the registered entity[18]. Further examples include registration in the aviation, healthcare, finance, and food industries.

In these industries, registries are a lever to increase regulatory visibility, support legal action and manage societal risks. Comparatively, frontier models exist in an abnormally undeveloped regulatory environment.

## What risks do frontier AI models pose?

Experts have identified many risks and harms that current or near-future AI could be capable of causing. Due to the flexible nature of frontier AI models today, these risks span many domains, including but not limited to:

- Economic disruption, through large-scale automation, delegation, or reorganization[19] of jobs across many skill levels and domains[20,21], and the introduction of many many automated economic agents.
- Cybersecurity, by lowering the barriers to entry for launching sophisticated or automated cyber attacks[22,23].
- Biosecurity, terrorism, and nuclear non-proliferation, by lowering the barriers to entry for developing biological[24] or chemical[25] agents[26], and even radiological or nuclear weapons[27].
- Undermining of democratic values, by enabling far-reaching disinformation or deliberately manipulating people, undermining political institutions[28] and increasing international tensions[29].

These and other risks[30] could be exacerbated if, as some experts predict, frontier AI becomes harder to control and harder to align as their ever-increasing capabilities outpace safety research and governance[31]. AI labs recognize these risks and are working to identify and reduce the chance of harm. For example, they are working on research into measuring the capacity of individual models to autonomously self-replicate[32], to deceive humans, or to enact long-term plans that demonstrate situational awareness[33]. However, there is widespread consensus that state of the art AI research is not yet effective enough to identify all possible risks from AI technologies, nor to mitigate them[34,35].

## What is an AI model registry?

A registry is a centralized database designed to collect, store, and manage information about particular products, services, technologies, and economic actors such as corporations and professionals. They're used by many governments, regulatory bodies, and other organizations to provide insight into services, products, and their manufacture and to track the legal identity of the people and entities responsible for possible harms. For example, the US Food and Drug Administration maintains registries of manufactured medical devices[36], food facilities[37], drugs[38], and more, which each contain information on the products themselves, their safety and risks, their manufacturers, and so on.

Registries let the government and public know what products or services are being sold, how they're made, and by whom. They act as a lever to increase regulatory visibility, support legal action and accountability, and manage societal risks.

To do so, they typically require individuals or organizations to submit specific data about the entity being registered, which may include identifying and contact information, legal responsibility, technical specifications, intended use, and so on. Each category of information is useful in different contexts, as we'll explore fully in *Part II - Design of a Model Registry*. This information is typically stored in a standardized format to allow for easy access and processing, though registries differ in how much access is granted to the public, approved government administrators, and third parties, as we'll discuss in *Should the information in the registry be confidential?*. Throughout this report, while discussing the value and downsides of individual design features and information categories for an AI model registry, we'll refer to registries from other industries to provide context and comparison.

Registries also play an important role in broader regulation and oversight. They enable authorities to monitor compliance with laws and regulations, identify trends or potential issues within a particular domain, and develop high-quality, evidence-based policy. This is further supported by the communication between regulators and registrants that a registry can provide, serving as a channel for updates, notifications, and ongoing reporting requirements.

In the rest of this article, we'll advocate and recommend design decisions for an *AI model* registry. That is, a registry of AI models, containing information on their development, capabilities, risks, details of their deployment, responsible parties and so on. We believe that such a model registry would be a powerful and foundational tool in the governance of AI.

## What are the specific benefits of creating an AI model registry?

An AI model registry should be a critical component of an effective AI governance strategy. It would lay the foundation for understanding the state of frontier AI development, permit regulatory enforcement on AI developers, enable further legislation, and foster public sector field-building. We describe in detail the purposes and benefits of an AI model registry below:

### 1. A registry facilitates the monitoring of frontier AI technology, including model capabilities & risks

Governments currently lack basic insight into frontier AI models, as we've discussed in *What value does a model registry provide to governments?*. Consequently, they also lack the capacity to predict safety concerns and identify possible mitigations. A registry would provide governments with increased regulatory visibility into current frontier model capabilities, associated risks, and potential safety concerns. It would also enable more accurate forecasting around the future development of frontier models, and support predictive tools, such as AI scaling laws, to provide reliable estimates of when and where AI is likely to have particularly significant societal impacts[39]. This visibility would be foundational in identifying and managing existing and future risks.

An AI registry could also facilitate international responses to unforeseen incidents arising from AI systems. It would enable more effective global monitoring of AI development trends, and serve as a resource for coordinating global responses in the event of an unforeseen, extreme event. The registry would enable rapid information sharing across borders and coordinated mitigation efforts, similar to how national nuclear material registries support global non-proliferation efforts, or how national disease surveillance systems support global pandemic responses.

## 2. A registry provides a visibility for accountability & regulatory enforcement

A registry would allow authorities to track when qualified models meet any existing regulatory criteria, and verify their compliance. It would support informed, timely interventions in the event that a developer has not met any mandatory safety standards that may emerge in future. The visibility provided by a registry would support a shift from a trust-based model, where organizations are expected to comply voluntarily, to an ecosystem in which governments can ensure that developers are accountable to current and future legislation.

As an example, governments may mandate further safety evaluations[40,41] or 3rd-party audits, improved information security practices to mitigate the risk of misuse, or, in extreme cases, development pauses on the models in question[42]. In these cases, a registry would be essential for identifying non-compliant models, and mitigating associated risks. A registry would also be important for proactively identifying violations of legislation that already exists, such as restrictions on developing biological weapons or export controls[43,44].

Finally, by providing governments with visibility into AI capabilities and impacts, registries could enable lighter-touch interventions before issues escalate and require heavier regulation[45]. Including contact information in the registry, as is already implemented in the EU AI registry, would streamline communications between government and industry for routine correspondences, emergencies, and the examples mentioned above.

## 3. A registry enables the development of new regulation as necessary

The proposed model registry would enable more precise and targeted AI regulation by providing governments with a framework to effectively classify models based on specific capabilities or characteristics. This approach ensures that any potential policy is intelligently calibrated to empirical evidence about risks and effective mitigations, determining precisely if, when, and what regulation may be needed.

Importantly, this registry is not intended as a "foot in the door" for excessive regulation. Rather, it aims to equip policymakers with broader insights and crucial information to make informed decisions about AI governance. This data-driven approach allows for the design and enforcement of effective governance measures only if and when empirical evidence demonstrates a clear need.

Examples include:
- A model registry could serve as a regulatory hub for categorizing and

passing legislation on specific models based on characteristics submitted to the registry, such as capabilities (e.g. biological abilities), open-source status[46], or computational thresholds.

- A model registry could enable the enforcement of responsible scaling policies[47], by allowing the implementation of regulation that requires greater safety measures based on model capabilities or compute thresholds.

- A model registry could enable the implementation and enforcement of licensing systems based on model characteristics such as capabilities, use-cases, or compute thresholds.

- A model registry could be tightly integrated with a mandatory incident reporting[48] database, allowing incidents to be mapped to existing AI models.

- Mandatory third-party evaluations[49] regarding safety and alignment could eventually be integrated into a model registry.

- A model registry could enable the design and enforcement of *tiered access controls* for AI models based on their capabilities and potential risks. For example, a future governance policy may restrict access to the most powerful models while allowing broader access to less capable or lower-risk models.

- A model registry could enable the development and legislation of mandatory on-off switches for AI models above a certain threshold of capability[50].

- A model registry could enable the enforcement of mandatory impact assessments. Impact assessments are a fundamental part of risk mitigation – for example, the EU AI Act requires that high-risk AI systems undergo conformity assessments. Governments could require that certain AI models submit impact assessments to provide safety assurances[51,52].

- A model registry could enable governments to eventually audit specific AI labs or models based on reported model features or risk factors, allowing for more direct oversight of compliance with safety standards.

## 4. A registry fosters public sector technical and regulatory expertise

A registry could promote the use of common standards and best practices across government AI regulations. It could create standards for reporting and categorizing AI models, leading to shared language and characterisations of different systems and their associated risks. This would also raise awareness of AI impacts to a broader range of policymakers[53].

There are many public stakeholders who do not work directly on AI but whose decisions will be influenced by, or influence the far reaching impacts of frontier models. By providing structured information on AI risks and potential mitigations, the registry would help policymakers across various domains understand AI's relevance to their work. Working from the same source of information, different agencies would be more likely to align their regulatory approaches to AI-related issues.

Implementing model registries would also cultivate AI expertise in the public sector[54] by encouraging the development of teams with the technical skills and knowledge required to manage the registry. Specialists would develop skills in understanding, interpreting and critically assessing model capabilities & safety evaluations. In addition to addressing the current concentration of expertise solely in industry, this required technical knowledge would likely have other ancillary benefits across government[55].

## What risks are associated with creating an AI model registry?

While registries are a relatively lightweight and low-risk tool to increase the capacity and technical awareness of governments, they also have downsides, including the following. Our proposal for a model registry is intended to mitigate these downsides.

### 1. Registries can contain sensitive information.

Proprietary information or intellectual property such as model design, training algorithms, and sources of training data can represent commercial advantages that AI developers would want to protect. Including such information in a public or insecure registry would impact the competitive landscape for AI developers and result in significant pushback.

Some information can also be hazardous. For example, techniques used to train frontier AI models could be misused to develop a model independently by parties that lack the knowledge and tools to ensure safe development. Sharing data could encourage race dynamics, which safety advocates are keen to avoid[56,57].

### 2. Excessive reporting requirements could burden both AI labs and registry administrators.

A model registry that requires too frequent or too detailed updates may slow the pace of innovation and AI development, or in extreme cases deter compliance. Additionally, such a registry may require significant resources from the administering body, which may not be available from a governmental budget.

### 3. A registry may not be accurate or useful unless it is enforced via consequences for non-compliance.

AI developers have meaningful incentives to avoid reporting to a registry, including time overhead and avoiding further regulation. To ensure compliance, developers will need to face consequences such as monetary fines. Additionally, it may be necessary to require that enterprise customers of AI models verify that the model they are using is registered meaning that developers face a market penalty in loss of sales if they are unregistered. Later in this report, we'll discuss a plausible implementation of these incentive mechanisms.

## What AI model registries currently exist?

This is not the first call for a framework for registering key information on AI models. Some AI labs voluntarily share data and safety information with governments[58], directly and through standards such as model cards[59]. The US, EU, and PRC have started to develop mandatory AI registration, though these do not include requirements for rich information sharing, and each serve different specific governance functions.

The first significant model registry was established in New York city by a 2019 mayoral executive order[60] and covers governmental algorithmic tools that are derived from complex data analysis; support agency decision-making; and have a material public effect[61]. Entries included the agency, name of the tool, the date it entered usage, its purpose, and its overall function, though later the NYC AI Action plan[62] lead to richer information, including details of training data, type of model, and whether identifying information was stored.

China announced the their national AI registry in 2021 in their Algorithmic Recommendation Provisions[63], which has been expanded by further provisions in 2022 and 2023. Developers of algorithms that synthetically generate novel content, or which display "public opinion properties" or "social mobilization capabilities", must report basic data such as the provider's name, domain of application, and a self-assessment report to an algorithm registry within 10 days of publication[64]. The registry is designed primarily to manage the impact of generative AI on public opinion and prevent societal disruption[65].

The EU AI Act, published in May 2024 and entering force in August 2024, requires systems that have been classified as high-risk - determined by their use case and including AI used in critical infrastructure, education, law enforcement, and migration[66] - to register basic contact information[67]. This primarily serves as a contact directory, while other parts of the act impose separate safety requirements on high-risk models, such as risk assessments, training data standards, activity logging, and robust security requirements.

In the US, President Biden's 2023 Executive Order on AI requires developers to notify the government of all models trained with computing power above a threshold of $10^{26}$ floating point operations, though no existing models qualify at the time of publication. Developers of qualifying models that also demonstrate dual-use capabilities will be required to file reports with the government, including reporting[68]:

- Ongoing or planned activities related to training, developing, or producing dual-use foundation models, including the physical and cybersecurity protections taken to assure the integrity of that training process against sophisticated threats;

- Information on model weight ownership and corresponding security protocols;
- Outcomes of any relevant red-teaming tests[69].

The order also calls for NIST to develop evaluation standards to guide the procurement and reporting of this information.

These registration efforts are nascent and may serve as a good foundation, but we believe they are not yet sufficient to achieve the full range of possible benefits from a model registry. They don't yet provide adequate insight for governments into the capabilities, architecture, compute used, and security of frontier AI models. We believe that additional, lightweight requirements to these registries could provide governments with significant regulatory advantages at little cost.

# Part II – Design of a Model Registry

In this section, we explore how to practically design, administer, and enforce a registry to have the greatest benefit for the least cost and risk. This includes discussions of what information should be required by the registry, as well as the consequences developers should face for non-compliance, how similar models can be registered jointly, and other design decisions.

## What design principles will minimize the regulatory burden of a registry?

Stricter regulatory requirements can hinder the speed of innovation[70] and lead to strong pushback from industry, as has already been demonstrated by AI labs funding hefty lobbying efforts[71,72]. To ensure a model registry fosters AI innovation while minimizing industry resistance, we adopt three principles throughout this report that aim to reduce the regulatory burden associated with such a system.

### Minimal Reporting

Overly stringent reporting requirements would needlessly divert resources away from research and development, create administrative obstacles to developing and testing new models. The burden of reaching compliance before deploying new AI systems would slow down the pace at which innovations reach the market.

Furthermore, a registry requiring labs to report proprietary information unnecessarily will create new concerns for the labs in question around intellectual property and the risk of leaking sensitive data (see [Model Security](#)).

Overly frequent, detailed or sensitive reporting requirements would also create a higher burden on the administrators of the registry itself and could be a barrier to its adoption and implementation. A registry must set a clear scope so as to only include those models that are required to meet its specific governance objectives, and not place burdens on labs unnecessarily.

Future reporting requirements should also be sensitive to the impact of excessive requirements on startups[73]. Currently only a handful of leading labs are capable of developing frontier models, however, in future, reporting requirements may also apply to start-ups or smaller organizations. In this case, regulators should take care not to overburden small organizations beyond their capacity.

### Our Recommendations

**We recommend that AI model registries should aim to *only* require the information that is needed for the above purposes.**

- A registry should carefully consider what information is essential to meet its governance goals (see *The Case for a Model Registry*), and exclude information that cannot be clearly linked to achieving those goals.

## No Mandatory Standards or Licensing

Many industries enforce safety measures or protect the rights of users using mandatory standards or licensing schemes. Mandatory standards are rules or requirements that organizations in specific industries must comply with. These standards are enforceable, but enforcement may take place after a product is deployed, and may not entail ever removing the product from legal use. Examples span many industries, including GDPR requirements, and standards in food safety, automobiles, aviation, healthcare, construction and finance.

Unlike mandatory standards which set rules for how activities should be conducted, licensing schemes determine who can engage in certain activities in the first place. Licensing schemes take effect before a product is released onto the market, and involve mechanisms for removing the product from the market if it fails to meet the requirements of the license. Examples include licenses to practice medicine or law, and pharmaceutical manufacturing licenses for drug production.

Both licensing and mandatory standards are effective governance tools in mature industries, where relevant risks mitigations are well understood, and can be enforced with confidence. However, AI safety evaluations, lab cybersecurity, incident reporting, and general safety standards are still immature. Because standards and best practices are still being developed in these domains, the key function of a model registry today should be to function as a lightweight tracker to build capacity and information gathering for governments, providing a foundation for robust policy and standards in future.

Licensing and mandatory standards generally require more resources from both regulators and companies compared to maintaining a registry, and enforce a much higher regulatory burden on AI labs. Combining an early registry with licensing requirements is likely to slow the deployment of models to the public and face blowback from AI labs while at the same time enforcing standards and behavior whose implications would not be well-understood. A registry may eventually inform the development and enforcement of mandatory standards or licensing schemes, but the registry itself should be considered as a distinct governance tool.

### Our Recommendations

**We recommend that model registries should not include mandatory standards or licensing requirements.**

- A registry is primarily intended to support future regulation, rather than as a stand-alone regulation itself.
- It should be designed for reporting existing information about an AI model, and require minimal additional overhead for developers.
- A model registry may be a good foundational platform for creating licensing requirements separately, if they are found to be prudent in future.

## Global Interoperability

Registration requirements may create additional challenges for labs operating across multiple countries. Different national registries that are not interoperable and/or require different information to be reported may lead to duplicated efforts and inefficiencies as companies navigate disparate regulatory frameworks.

In order to minimize the overhead of labs operating across jurisdictions, a registry should be developed with careful consideration of existing national reporting requirements for AI. The registry should avoid reinventing the wheel where adequate standards already exist, and otherwise take measures to align with equivalent systems in other jurisdictions. (See *What AI Model Registries Currently Exist*). Ideally, reporting requirements of different national registries would be sufficiently aligned so that registering with several would not create a much greater administrative burden than registering with one.

Beyond aiming to synergise specific reporting requirements, registries could also be interoperable through mutual recognition: A registration in one country could be recognised by another, and meet the reporting requirements for both jurisdictions. Eventually, different nations may adopt the same international registry so that a single registration would apply across many different nations, as the EU AI Act registry aims to do.

Relevant standards around cybersecurity (*Model Security*), model evaluations (*Model Evaluations*) and others will likely change over time, and registries should be prepared to update reporting requirements as new information comes to light. This entails both monitoring the state of international standards, and developing mechanisms to update registration requirements smoothly.

### Our Recommendations

**We recommend that an AI model registry should be interoperable with international standards.**

- Interoperable standards would be particularly valuable for emerging topics such as the cybersecurity of AI models, model evaluations, and post-deployment monitoring.
- Few standards currently exist for these domains. We recommend that governments fund the development of improved international standards, and continually track and implement the most effective and universal standards as they are developed. This could include:
  - Implementing a regular review process (e.g. annual or biennial) to assess and incorporate new standards.
  - Initiating international dialogues with the aim of aligning approaches to reporting requirements and registration[74].

## What should qualify for inclusion on the registry?

### What thresholds should a model exceed to qualify for inclusion?

Our goal is to have a registry of all AI models that pose a potential risk of significant harm or disruption, but we don't want to bloat the registry with thousands of low-risk AI models. To find the right balance, we need to identify criteria for including models in our registry.

The bar for inclusion needs to be low enough to give the government and policymakers insight into models before they present significant risks, in order to lay the foundation for sufficient information collection and risk mitigation. Since major harm has not yet emerged from existing frontier models, one promising option would be to set inclusion criteria just above today's most capable models. Such a threshold for inclusion would ensure comprehensive coverage of the next generation of AI models, while minimizing retroactive overhead for existing models.

Experts are most concerned about harm from the most powerful AI models and from models trained for use-cases in high-risk fields[75] such as nuclear security[76] and biotechnology[77]. As discussed in the previous section, most governmental efforts[78] to narrow the scope of legislation to the highest-risk AI focus on use cases and capabilities.

Robust and sensitive capability evaluations would be ideal for this, and future AI governance will likely rely on capability evaluations as the most accurate way to determine risk. However, capability evals are currently inadequate for this task[79] and capability benchmarks for existing generations of frontier AI have become obsolete[80]. Better evaluations are under rapid development (as we'll discuss in _Model Evaluations_), but it's unclear when they will catch up with the rapidly advancing frontier of AI capabilities. In light of this uncertainty, a model registry would benefit from tracking variables which provide direct proxies for overall capabilities, such as the following[81]:

**The size of a model is a useful proxy for general model capabilities.**

For an introduction to model size, see _Model Size & Parameters_.

In summary, the size of a model, measured in total number of parameters, or in average number of active parameters per token, is a useful proxy for capability.

**Compute is a useful proxy for general model capabilities.**

For an introduction to "compute", see _Compute Used For Training_.

In summary, computing power or "compute" typically refers to the amount of computational resources required to train a model, measured in floating point operations. Compute is a popular target for governance[82] as it has a direct impact[83] on the resulting capabilities of the model.

**Amount of training data is a useful proxy for general model capabilities.**

For an introduction to training data, see _Training Data_.

In summary, the amount of training data used to train a model, measured in number of tokens, is a less common but still useful proxy for capability.

Note that the size of a model, compute, and training data are all closely linked in how they affect model capabilities, in an area of research called scaling laws (see _Key concept: Scaling laws_). In short, it's sensible to track all these variables in a registry, and to use distinct inclusion criteria for each.

**High-risk domains**

While overall capability is a useful proxy for risk, we should be especially cautious of AI trained for, or deployed in, certain high-risk fields[84]. AI systems deployed in high-risk fields can pose great risks even if they don't cross our thresholds for compute, training data, or size. Furthermore, specialized AI models can be much smaller and less general while still exhibiting high-risk capabilities[85].

High-risk domains include:

- **Nuclear power and weaponry**
    - AI involved in the maintenance and storage of nuclear weapons or

nuclear waste, in the chain-of-command of using nuclear weapons[86], or in the functioning of nuclear power plants[87], would be high-risk. This is due to the catastrophic impact of failure, whether by cyberattack, misalignment, or AI malfunction.

- **Chemical weapons and biological weapons, pharmacology, synthetic biology, and biological design tools**
    - AI that is demonstrated to lower the barrier of entry[88] to developing chemical[89] or biological weapons[90] could generate catastrophic harm.
    - Some biology research involves the production and modification of genetic material, including the remote production of custom-generated DNA and RNA molecules[91]. This poses a particular risk, as a malicious actor or AI could use these systems to generate a pandemic-causing virus[92].
    - Generative AI has also been used in biological design tools[93] to predict protein structure, providing unprecedented capabilities to design proteins for custom tasks. Malicious actors could use these biological design tools to create catastrophically harmful biological weapons.
- **Cybersecurity**
    - AI that is highly capable in the domain of cybersecurity[94] and cyberattacks[95] could bypass many safety features and cause catastrophic harm by granting access to hazardous information or crucial infrastructure that could be remotely sabotaged.
- **Self-improvement and autonomous replication & adaptation (ARA)**
    - These refer to an AI model's ability to independently propagate, adjust to new situations, and enhance its capabilities. This includes acquiring resources, obtaining more computing power, installing itself on new systems, self-improvement, and adapting to challenges. An AI capable of ARA[96], as these skills are collectively termed, could create numerous self-improving copies, leading to rapid and unpredictable growth in its capabilities and influence.

To account for risks from specialized AI systems, governments would ideally prefer to use capability evaluations that demonstrate dangerous capabilities in each of these domains. Without effective capability evaluations, governments may prefer to choose custom inclusion thresholds for compute, training data, and model size for AI models that have training data in such fields, as the US Executive Order does for AI used in biotechnology[97].

To account for risks of AI systems being deployed in high-risk domains, governments should aim to comprehensively identify a list of high-risk domains, and mandate specific additional safety requirements for AI used in those domains. This is the approach taken in the EU AI Act, which classes systems used in critical infrastructure, education, law enforcement, and other domains as 'high risk systems', which are subject to additional restrictions.

## Our Recommendations

In light of these tradeoffs, **we propose that AI models should initially qualify for inclusion on a registry if they exceed any of a set of thresholds relating to:**

- Compute power used during training, measured in floating point operations;
- Amount of data used during training, measured in number of tokens;
- Model size, measured in total number of parameters and average number of active parameters per token;
- Specific high-risk capabilities or training regimes, using custom inclusion thresholds, assessment of data sources, and, where tenable, capability evaluations.

In the future, when capability and risk evaluations are improved, these proxies should be replaced or augmented to include such evaluations.

We provide an example of numerical values that exclude all AI models deployed before Jan 2024, but should capture the next generation of frontier AI models. For each value, we will cite evidence that this will not include existing models and, where possible, we'll cite examples of other registry proposals that use similar values. In this example, models should be registered if they meet any of the following conditions:

- They were trained with at least $10^{26}$ floating-point operations[98,99];
- They were trained with at least $10^{14}$ tokens of training data[100];
- They use at least $10^{12}$ active parameters while running[101];
- They were trained with at least $10^{23}$ floating-point operations[102] and were trained primarily with data relating to, or have been demonstrated to be capable of lowering the barrier of entry to, any of the following high-risk areas[103]:
    - Nuclear and radiological technology or weaponry;
    - Chemical weapons or the effect of chemicals on humans;
    - Biological sequence data, biological weapons, or biological design tools;
    - Cybersecurity or cyber-attacks;
    - Autonomous replication, adaptation, and self-improvement.

---

[98] Notable AI Models - Epoch Note that the model that used the most training computation at time of publication is Gemini 1.0 Ultra, using an estimated $10^{25}$ floating point operations. See also Computation used to train notable artificial intelligence systems - Our World in Data for an alternative presentation of the same data.

[100] Notable AI Models - Epoch; Note that the model that used the most training computation at time of publication is Llama 3.1-405B, using an estimated $1.6 \times 10^{13}$ tokens. See also Datapoints used to train notable artificial intelligence systems for an alternative presentation of the same data.

[101] Note that we lack access to the number of active parameters of some models. According to Notable AI Models - Epoch, the model with the largest absolute total parameters is "QMoE: compressed 1T model", a mixture-of-exports model with $1.6 \times 10^{12}$ parameters. Since only a small minority of these are likely active during use, $10^{12}$ active parameters is likely to be higher than any current model.

[102] This number is based on the Executive Order on the Safe, Secure, and Trustworthy Development and Use of Artificial Intelligence - Section 4.2.b.i., which lowers the threshold from $10^{26}$ to $10^{23}$ for models "using primarily biological sequence data".

[103] This number is based on the Executive Order on the Safe, Secure, and Trustworthy Development and Use of Artificial Intelligence - Section 4.2.b.i., which lowers the threshold from $10^{26}$ to $10^{23}$ for models "using primarily biological sequence data".

> This list of high-risk areas could be adapted to the relevant national security concerns of the government enforcing, but are broadly taken from similar high-risk areas cited in the EU AI Act[104] and the US Executive Order on AI[105] and METR's focus on autonomous replication[106].
>
> Note that increasing algorithmic efficiency means that models in the future will be more capable with the same amount of compute, data, and number of parameters[107]. Due to this and uncertainty around the risk posed by next-generation frontier models, these inclusion criteria will need to be continually modified to maintain a useful threshold for identifying risk.

### At what stage would a system be required to be registered?

AI development is a complex process with many stages before and after deployment. Key stages include data preparation, model development, model training, validation, fine tuning, and testing[108]. A registry could require developers of AI models to register at any point throughout this timeline.

Early registration would naturally grant regulators early insight into AI development with plenty of time to analyze and process the provided information before the model is deployed. However, registering early in this process will make the information less reliable – developers won't be able to provide assessments of capabilities or the amount of compute used during fine-tuning early on, for example. Further, early registration will increase the regulatory burden on both the registry and the developers, as models will undergo many rounds of development without ever being deployed to the public, as was the case with models in the GPT-3 family[109]. This will likely be frustrating for AI labs that are experimenting with new models and designs or that don't plan to deploy their models publicly.

Different industries face different requirements on the timing of registration. For example, the FDA requires registration of new drugs before clinical trials begin[110]. Pesticides must be registered with the EPA before manufacturing begins[111]. Nuclear facilities must be registered and undergo reviews from the design phase onwards[112]. Though there are significant differences, in all these cases the key is that the models must be registered before they have the capacity to do harm to the public.

It is certainly plausible for models to cause harm before deployment, through loss of control during testing, leaking or exfiltration of details of the model architecture or its weights, and so on. In the long-term, governments should adopt policies that give regulators and safety experts insight into models during development, matching registries of clinical trials, pesticide manufacture, and nuclear facility design listed above. However, this would require much more significant government infrastructure and cooperation from AI developers, compared to registering models only before public deployment. Further, public deployment significantly expands the potential for harm by exposing AI models to a broad range of users not employed by AI labs. It increases a model's attack surface and enables external parties to exploit model capabilities in ways that may not have been anticipated by the developers.

### Our Recommendations

We recommend that models must be registered before they are publicly deployed. That is, in order to legally grant members of the public access to the model, through APIs or directly, the developers of that model must register the model. This ensures that registration occurs before the attack surface of models is substantially increased, ensures the information is accurate at the time of deployment, and lightens the regulatory burden on both developers and regulators.

## What constitutes a new model, or a model update? When should re-registration occur?

The distinction between one model and another can be unclear as similar underlying models can be fine-tuned[113], re-trained[114], iterated[115], presented within a new user-friendly interface[116], or given different computational or data resources. For example, upon publication OpenAI's GPT-3 was actually a family of 8 similar models that varied in amount and source of training data; in size over three orders of magnitude; and which were fine-tuned for different purposes[117]. Further, only some of this family of models were ever publicly deployed through APIs and the ChatGPT UI. Today, the set of models that could be categorized under the heading "GPT-3" has grown to at least fifty[118].

Requiring developers to submit new entries to the registry before they can deploy each iteration of a model would create undue burden for both the developers and the administrators of the register. However, we don't want information in the registry to become inaccurate or outdated.

Registries in other industries resolve analogous issues by allowing single entries to represent families of products or services. For example, the FDA allows similar medications[119] - those with the same active ingredient, differing only in dose strength - to be registered singly, while drugs with different delivery mechanisms or otherwise differing pharmacokinetics must be registered independently. We could apply this approach to AI models by allowing multiple similar models to be represented by a single "family" of models.

However, as demonstrated above with the populous GPT-3 family, the types of "model families" that developers use spread across many orders of magnitude across multiple parameters (size, training compute, training data, etc), and different developers may categorize their models differently. Further, a new model with multiple orders of magnitude more compute power could have drastically greater capabilities and could plausibly cross important thresholds, for example by enabling recursive self-improvement[120].

A solution would be to allow a model family to be represented by their most capable member in each key measurable dimension, since these iterations are the one that governments are most likely to be interested in. Additionally, all models within a model family would be held to the reporting standards of its most capable members.

Finally, the registry should be designed to ensure entries remain accurate over multiple years. Developers could potentially lower the cybersecurity or provide new API access to older models, years after deployment. While this might not trigger a new entry in the registry, this information is still important to policymakers and regulators. Therefore, to maintain an accurate registry, entries in the registry should be updated on a regular basis.

The US BIS released a document in September 2024 detailing prospective frontier model reporting requirements that must be updated on a quarterly basis, including any qualifying activities undertaken in the given quarter, and projected in the next 6 months[121]. However, unconditional quarterly requirements are likely to create constant administrative pressure on labs, and are not likely to capture any additional safety-relevant information that would not already be captured by measures discussed above and less frequent updates. However, updating registry entries in intervals of a year or longer will likely lead to the registry being outdated towards the end of the reporting period given the rapid progress of frontier AI labs[122].

### Our Recommendations

We recommend that developers may be allowed to add and update qualifying models as a new *model version* as part of an existing *model family*. Such a model family would primarily record the entire set of reporting requirements for the *most capable model version* along each key measurable dimension.

We recommend that developers only be required to submit a *complete registry submission* for a model version in the case that the new model exceeds the most capable model versions in its family by a certain amount (e.g. 20% of model size), or released some time interval after initially registered (e.g. after 3 years of initial registration). If it does not exceed the most capable model versions and was deployed within the time interval of the initial registration, it will be sufficient to simply report the name of the new version for tracking purposes.

Ideally, as we discuss in *Model Evaluations and Risk Assessments*, these key measurable dimensions would be based on capability assessments. As we currently lack good assessments of capability, we recommend currently using the following key measurable dimensions as proxies for the "most capable" model:

1. Model size
2. Total compute used during training and retraining
3. Amount of training data
4. Specific powerful capabilities, such as the ability to generate CBRN infohazards, generate deepfakes, conduct autonomous replication, or improve cyberwarfare abilities.

For the first three key measurable dimensions described, a new submission to the model registry would be required when a new version exceeds the previously most capable model by 20%. For the measurable dimension of "specific powerful capabilities", a new submission would be required when a new version exceeds the previous most capable version in its results from a related capability evaluation by a certain amount (e.g. 20%), or when crossing a certain threshold score (e.g. scoring 80% in an autonomous replication capability eval). Note, however, that reliable evaluations of this sort do not yet exist.

We also propose that models must be registered in a new family if they're deployed more than 2 years after initially registered. This is important, as algorithmic progress means that models in 2030 could be far more capable than models in 2027 while still the same size and trained with the same compute and training data[123].

To avoid undue pressures on labs while ensuring information is up-to-date, developers should be required to update their register entries twice annually to ensure the information is accurate, regardless of updates to the model family.

This system would significantly reduce the overhead of reporting updates for developers, by concentrating registration requirements solely on the most capable model versions. New model development could occur without necessitating re-registration, as long as they do not meaningfully exceed the most capable version along a key measurable dimension.

---

[122] Examples include the EU AI Act's high-risk AI system register, Helsinki's AI Register, the FDA's various registers such as their medical device register, and many national company registers.

> One issue is that organizations will be incentivized to submit all their new models in a single model family, to minimize the amount of reporting. However, registry administrators would prefer that a model family represents a group of meaningfully similar models. To prevent model developers from simply submitting all new models into a single family, we recommend that all models in a family must meet similar reporting criteria and binding requirements in future legislation.
>
> As an example, we recommend that all models within a family must meet the same submitted security requirements and open-source status. Similarly, if a version in a model family crosses a future threshold for a new requirement (e.g. requiring increased cybersecurity due to its capabilities), that requirement will hold for all versions in the model family.
>
> This will incentivize developers to categorize their models into appropriate model families according to their intended use-cases and capabilities, to minimize the requirements with which their models must comply.

## What information should an AI registry contain?

### Basic Information

Most registries include basic information about the organization that produces or distributes the registered product or service[124], such as the organization's name, corporate or charitable status, contact information, senior management and authorized representatives, and sometimes more involved information such as affiliations and sources of funding. This allows regulators to identify responsible parties in case of incidents, assess potential conflicts of interest or undue influence, contact developers quickly and directly, and ensure compliance with regulations.

This information is generally low-risk to include in a registry, as much of it is publicly accessible through disclosures and standard company and non-profit registration. However, some information may still be sensitive and AI developers may not favor sharing detailed financial information, client lists, or the personal information of individuals. Registries can balance these tradeoffs by securing more sensitive information.

## Our Recommendations

We recommend that developers of a qualifying AI model must have a legally recognizable entity. We recommend the following basic organizational information should be required from that entity:

1. Legal business name of the developer/owner, and any trade names or aliases;
2. AI model family trade name;
3. Unique name for each AI model version submitted;
4. Status of an AI model version (i.e. on the market, recalled/no longer on the market, etc);
   a. Date of model deployment (if deployed);
5. Business registration number or tax identification number;
6. Legal structure (e.g., corporation, LLC, partnership);
7. Registered address and principal place of business;
8. Contact information, including phone number, email address, and emergency contact details;

This list isn't exhaustive, and a model registry could require additional information that may be less vital or practical to share, such as:

1. Names and titles of key individuals (e.g. CEO, CFO, CTO);
2. Board of directors or governing body members;
3. Annual revenue from a model;
4. Regulatory licenses or certifications held;
5. Insurance information (e.g. liability coverage);
6. Number of employees;
7. Ownership structure;
8. Parent company or subsidiaries (if applicable);
9. Stock exchange listing (if public);
10. Foreign ownership percentage;
11. Cybersecurity certifications;
12. Major model clients or government contracts;

## Open-Source Status

Some software developers *open-source* their software, freely sharing part or all the underlying code and data and allowing anyone to use their work and products. The range of definitions of what constitutes an "open-source" AI model is larger than for traditional software, as AI developers can choose to share many components of the model, such as[125]:

- Sharing the model weights - the parameters that determine the model's capabilities, set during training. These weights allow a model to be used and fine-tuned without expensive training or development;
- Sharing the complete set of training data used to create the model;
- Sharing the underlying source code and architecture.

For example, Meta released the weights of Llama-3[126] but not their training code, methodology, data, or model architecture. This is often called *open-weights*[127] to distinguish from total open-sourcing, though there isn't yet a consensus on what an "open-source" model precisely refers to[128,129].

Regulators need to know how open each qualifying model is - precisely which components of the model are open-source, if any - in order to design and enforce effective safety-focused AI regulation. This is because open-source models pose particular threats that closed-source models don't, likely requiring stronger and more targeted regulation. **Designing and enforcing such regulation will require insight into which models have open-source properties and to what extent.** For the purpose of this discussion, we will treat "openness" as a spectrum, and use "open models" to refer to models that have any combination of the above open-source components.

- **Open models cannot be controlled after they're deployed. Any effective regulation will need to verify the safety and alignment of open models before deployment and to higher standards.**
    - If vulnerabilities or dangerous capabilities are found in a closed-source model, the people deploying that model can deny access until the hazard is fixed. In contrast, if an equally capable but open model was found to be dangerous, there's no single point of access that could be denied[130]; other actors may have already reproduced the model, hazards and all.
    - Therefore, open models may face stricter requirements in demonstrating safety and alignment in the future[131], making open-source status a useful statistic to track.
- **Open models are far easier to replicate and misuse, and future regulation will be needed to assess and reduce their misuseability.**
    - Open-sourcing frontier models leads to greater proliferation of these powerful tools, increasing the risk of accidental and deliberate misuse[132], and lowering the barrier of entry to certain high-risk domains, such as bioweaponry[133].
    - Open models are easier to alter and therefore more vulnerable to misuse. It's much easier to undo the fine-tuning[134] of open-weights models (a step during development that improves the reliability and

[142] You can read a more detailed summary on the benefits and risks of open-sourcing AI here.

accuracy of model outputs). This was experimentally demonstrated with BadLlama[135], built using the open-weight Llama-2, in which fine-tuning was easily removed. One author wrote: *"You can train away the harmlessness...with currently known techniques, [but] if you release the model weights there is no way to keep people from accessing the full dangerous capabilities of your model with a little fine tuning."*[136]

- Any regulation designed to reduce misuse will need to have separate constraints for open-source models, and such regulation will be harder to design and enforce if regulators don't have a clear picture of which and to what extent models are open-source.

- **Open models spread AI development techniques, leading to increasing AI capabilities.**
    - Sharing cutting-edge training and development techniques from frontier AI labs will lead to a faster increase in AI capabilities[137], worsening an already large gap between AI capabilities and safety research/governance.

Proponents of open-sourcing models argue that these risks are outweighed, as:

- **Open models are easier to evaluate.** People have greater access to open-source and open-weights models, which allows third parties to identify biases or hazards in the model[138].

- **Open models give regulators more information.** Insight into model weights and underlying architecture gives regulators more information to make informed decisions and confirm compliance[139].

- **Open-sourcing resists the concentration of power.** AI labs are small and few but may have immense societal impact. The more information about models is shared, the greater the pool of people who have access to and influence over these powerful models[140].

There is neither expert consensus[141] nor public consensus[142] on whether the tradeoffs of open-sourcing mean that open models should face heavier or lighter regulation than closed-source models, and this disagreement is likely to persist as the AI landscape evolves. The EU AI Act treats open-source models favorably[143], exempting them from many obligations faced by commercial competitors unless the software is part of a general-purpose or high-risk system[144].

However, it seems likely that open models will face heavier regulatory requirements in other jurisdictions or legislation. As elaborated above, open-source models are harder to control once deployed, easier to replicate and misuse, and increase the spread of capability-enhancing techniques. These risks may far outweigh the benefits of open-sourcing if open models aren't under additional scrutiny before deployment. The EU AI Act's approach of relieving developers of open-source models of regulatory obligations may be increasingly dangerous as AI capabilities increase in the coming years.

Whether governments choose to more strictly regulate open models or not, the

[145] Note that these are distinct from hyperparameters, which are variables that dictate how learning is done.

**details of licensing and openness will be essential for designing and implementing future regulation**, and therefore the open-source status of a model is critical information to include on a registry.

### Our Recommendations

We recommend that an AI model registry should require developers to submit information on the licensing and openness of their models and components, including the following questions:

- What license has the model been released under? In particular, what rights and access do the public have regarding: copying, modifying, distributing, and sublicensing the model?
- Do users have access to the weights of the model?
- Do users have access to the training data used to train the model?
- Do users have access to the source code of the model or algorithms used to train or fine-tune the model?
- Are there any sub-components of the model for which the answers to the previous questions is yes? If so, provide details.

If users don't have access to the weights, data, or source code of any significant component of the model, it should be classed as closed-source. If users have access to the weights, but not the source code of the model, it should be classed as open-weights. If users have access to the weights and source code of the model, it should be classed as open-source.

## Model Size & Parameters

**The size of a model is a critical piece of information as it relates to capabilities, compute power, training data, and more, through what are called scaling laws.** Most frontier AI models are neural networks[145], consisting of nodes and connections between those nodes. The strength and sensitivity of these nodes and connections determines how the network functions, and adjusting these values is what happens during training[146]. The total number of adjustable values is often called the *number of parameters* of the model[147], and this measure of the "size" of the model. This is a useful proxy for the overall capability of a model[148]. As a reference, models in the GPT-3 family range from 125 million to 175 billion parameters[149].

However, some models[150] use architectures that have many more parameters while only a small fraction are *active* during use. These *sparsely activated* models can be much larger, but are still outperformed by smaller, densely activated models[151]. Therefore, counting the number of parameters that are active during use may be a better measure of capability, and therefore risk, than the total number of parameters. This value can be calculated by taking the average number of active parameters when running the model on a wide range of inputs. We'll refer to this metric as "number of active parameters" in the rest of this report. Keep in mind, though, that this measure has not been widely studied. Further exploration may be required as the science of capability assessments progresses.

---

[151] Note that floating-point operations, FLOP, is a measurement of total compute resources without reference to time. This is distinct from the computational *speed* of a machine, which is measured in floating-point operations *per second*, abbreviated to FLOP/s, or, confusingly, FLOPs.

Researchers at DeepMind have also derived a metric called *effective parameter count*[152], which adjusts for different activation and routing architectures to judge multiple different architectures on the same scale. However, this hasn't been widely adopted or independently verified, and is more complicated to assess. Nonetheless, customized measurement techniques like this may be useful for registries in the next few years.

Note that the size of a model, compute, and training data are all closely linked in how they affect model capabilities, in an area of research called scaling laws (see [Key concept: Scaling laws](#)).

To account for measurement uncertainty, regulators should allow some small margin of error in reported parameters of a model.

### Our Recommendations

We recommend that an AI model registry should require a measurement of both the total number of parameters of each model and the average number of active parameters during use across a wide range of model inputs.

To account for measurement uncertainty, the total number of parameters for a model should be accurate to within 10% of the true value.

## Compute Used For Training

Computing power, or "compute", is the amount of computational resources required to train or run a model, though usually refers just to training. Compute, measured in floating-point operations[153], is a critical factor in AI development because it determines how well the model adapts to its training data, impacting the complexity, accuracy, and capabilities of the model[154]. The compute necessary to train frontier AI has large and specific hardware requirements, and is relatively simple for developers to measure or determine based on their hardware and the model's training time[155]. These factors have made compute a popular target for governance[156], and both the US Executive

Order on AI[157] and EU AI Act[158] use compute as a proxy for capability and risk, setting thresholds around $10^{26}$ floating point operations. This amount of compute has yet to be used for any current models as of mid-2024[159], but is likely to be surpassed[160] by the next generation of frontier AI models.

Notably, however, developments in algorithmic efficiency[161], or how efficiently the algorithms are able to use compute during training and operation, mean that the compute necessary to train a model to a certain standard is decreasing exponentially (halving every eight[162] to nine[163] months). This means that models will become increasingly efficient, allowing them over time to reach capability thresholds with less compute[164].

Though initial training has the largest impact on a model, models also go through phases of fine-tuning, re-training[165], and post-training improvement that significantly adjust their behavior and capabilities[166].

These typically require far less compute power than initial training, but can have disproportionately large impacts on the model's behavior[167,168]. For example, fine-tuning (in the form of Reinforcement Learning from Human Feedback, RLHF) is what takes a text completion model like the base GPT-3 into a conversational interlocutor like ChatGPT[169]. Further, the same underlying model can be adapted into different iterations by fine-tuning or re-training, resulting in different capabilities (such as GPT-3 and ChatGPT, BERT and its domain-specific variants, and so on).

Note that the size of a model, compute, and training data are all closely linked in how they affect model capabilities, in an area of research called scaling laws (see Key concept: Scaling laws).

### Our Recommendations

We recommend that an AI model registry should require a report of the amount of compute, measured in floating point operations used to train the model. This metric should include both the initial training costs, as well as any retraining, fine-tuning, or post-training costs for a deployed model.

To account for uncertainties and difficulties in perfect measurement, reported values should be accurate to within 10% of an independent assessment of the value.

## Training Data

Deep learning models such as LLMs require data to train on. This data provides the patterns the model is trained to recognize and predict, and more capable models require more training data[170]. The amount of training data is typically measured in tokens[171], referring to the smallest unit of data that's useful. For example, when training a language model, a token could refer to a word or letter; GPT-3 was trained with 499 billion tokens[172].

Like compute, the amount of data used during training has been growing exponentially[173], and while less commonly used as a proxy for capability, training data provides an additional lever during development, and therefore an additional lens for regulators.

While the sheer amount of data used during training may make it difficult[174] for developers to accurately describe the provenance and type of all training data, regulators are already requiring coarse-grained reports on training data, which aim for a balance between practicality, transparency, and regulatory utility. For example, the US Executive Order on AI[175] asks the federal government to ensure that the collection, use, and retention of data is lawful, secure, and mitigates privacy and confidentiality risks. Similarly, the EU AI Act[176] describes the requirement for reports on high-risk AI systems that must: describe training data sets in general; their provenance; how the data was obtained; how it was labeled; and how it was cleaned.

Note that the size of a model, compute, and training data are all closely linked in how they affect model capabilities, in an area of research called scaling laws (see *Key concept: Scaling laws*).

These huge collections of data are gathered from many diverse sources: some are publicly available for use and scrutiny, such as the widely used non-profit datasets from Common Crawl[177]; others are scraped from public sources but kept private, such as OpenAI's WebText which was reportedly gathered from Reddit comments and links[178]; and others are of undisclosed and unknown origin[179]. Note that these sources of data can include copyrighted material, even those under free use licenses such as Common Crawl.

Whatever their sources, most AI labs' training data is proprietary and an important source of competitive advantage[180]. This is because, as discussed in *Key concept: Scaling laws*, increasing the size of and compute afforded to an AI model requires proportional increases in training data to have continuing impact on capability[181]. Therefore, AI labs rarely share their training data sets, or even share detailed information about them and their provenance; when they do, the amount of data shared is far from enough to train a frontier model[182]. Further, AI labs are hesitant to share the detailed provenance of their datasets as this can expose them to legal liability and public criticism[183].

These datasets are often privately scraped and can contain copyrighted material[184] or private user data[185], leading many to call for greater transparency of the content and source of such data[186]. While governments may exert greater control over this use of protected data, at present these sources of data represent a major strategic competitive advantage over which labs will want to maintain control and privacy. Therefore, requiring detailed information about training datasets is likely to lead to major pushback from AI labs and delay or disrupt the establishment of a registry.

[188] Discussed in one of our more light-hearted posts, [Understanding Epoch's Direct Approach - Zershaaneh Qureshi & Elliot McKernon](#)

### Our Recommendations

We recommend that the *amount* and *types* of data used to train a model should be reported and stored in the registry. The amount should be measured by the number of tokens, and be accurate to within 5% of the correct value to account for difficulties in measurement. Developers should be required to register the type of data by selecting categories from a list, for example whether any of the following were used:

- Text
- Images, including subcategories such as:
    - Labeled images of people
- Audio, including subcategories such as:
    - Isolated audio of human voices
- Video
- Genetic, biological, or bioinformatics data
- Toxicity, volatility, etc of chemicals or biological products

We don't recommend that the registry initially requires developers to disclose the source or copyright status of their training data. However, this could be an avenue for further development, for example by a registry acting as a vehicle for tracking copyright status or provenance.

### Key concept: Scaling laws

The relationships between compute, training data, and size, and how these affect capability, are studied through scaling laws. Essentially, increasing any of these individual variables - compute, training data, and model size - will increase the capabilities of the model, but these increases are magnified if all three are increased in the right proportions.

Figuring out what these "right proportions" *are* is crucial to scaling AI models effectively and efficiently. Thus, this is a fertile area of research, with empirical analysis by teams at DeepMind[187], Epoch AI[188], OpenAI[189], and elsewhere. However, these laws aren't ironclad and are still under scrutiny; different developers try to optimize their models in different ways, and researchers at Epoch AI have found inconsistencies in DeepMind's influential Chinchilla scaling laws[190]. Therefore, while they are linked, and data on each variable can contextualize other variables, it's sensible to include all these variables in a registry, and to use distinct inclusion criteria for each (as described in *What should qualify for inclusion on the registry?*).

## Model Architecture

The architecture of a model refers to the underlying design of the machine learning system; how the components are organized, trained, how they interact, and so on. This is a particularly complex and technical aspect of AI regulation; many models share a coarse architecture but differ dramatically in precise architecture, and many frontier AI models are distinguished by cutting-edge design. For example, the following models are all neural networks[191] using attention mechanisms[192], but differ in the technical details of their architecture and function:

1. GPT-3 uses a decoder-only transformer architecture for text generation, with a single stack of 96 layers for autoregressive language modeling[193].

2. DALL-E 2 uses a two-stage transformer-based architecture, with CLIP for text-image translation and a diffusion model for image generation[194].

3. AlphaFold 2 uses a hybrid architecture combining attention mechanisms with specialized components processing biological sequence data and predicting 3D protein structures, with 48 transformer blocks and iterative refinement of up to 8 passes to improve predictions[195,196].

These technical differences are difficult to interpret for regulators, but also difficult to meaningfully simplify. These complexities make model architecture both important and challenging for regulation[197]. It's important because:

- **Model architecture provides context for interpreting other metrics like model size and compute requirements.** This information allows for more accurate risk assessments and comparisons between models. For example, a model with fewer parameters but a more efficient architecture might outperform a larger model.

- **Developments in model architecture can lead to dramatic shifts in capability.** Since more capable models pose greater risks, registry administrators should be extremely cautious when storing information that could be used to advance frontier models[198].

It's challenging because:

- **Model architecture is incredibly technical.** This makes it difficult for researchers to provide concise summaries that would be legible to non-experts, and also makes simplifying design features such as choosing from a dropdown list untenable. It also makes it harder for regulators to verify the accuracy and specificity of provided descriptions, which could incentivize developers to keep their architecture descriptions vague and high-level.

- **Model architecture evolves rapidly.** New architectures emerge frequently[199] and existing ones are often modified. For example, the shift from RNNs to transformers, and subsequent innovations like GPT's decoder-only approach or PaLM's pathways system, demonstrate how quickly the field changes. This constant evolution[200] makes it challenging to create standardized categories or comparisons, and such a design would necessitate frequent updates to the registry's classification system.

- **AI developers are likely to be extremely protective of cutting-edge architectural design due to commercial sensitivity.** Revealing such details could potentially allow competitors to replicate or improve upon their innovations, eroding their market position.

Despite these complexities, the EU AI Act does require developers to share descriptions of model architecture in some cases (though the depth of description required isn't clear):

- For general-purpose AI models, technical documentation must include (and downstream providers must be informed about) *"the architecture and number of parameters"*[201];

- For high-risk or general-purpose AI systems with systemic risk, technical documentation must include *"a description of the system architecture explaining how software components build on or feed into each other and integrate into the overall processing"*[202,203];

### Our Recommendations

We recommend that an AI model registry should require developers to submit a description of the model architecture that balances the need for regulatory insight with the protection of proprietary information and security concerns.

- Developers should provide a high-level technical description of the model architecture, sufficient for an expert in the field to distinguish it from similar models with different performance or functions.
- The description should include the general type of architecture (e.g., transformer, mixture-of-experts, etc.) and any significant innovations or departures from standard architectures.
- Developers should report the number of layers and the types of layers used (e.g., attention layers, feed-forward layers) without disclosing precise configurations.
- Developers should disclose if the model uses any form of external memory or knowledge retrieval systems.
- The description should include information on whether the model uses multi-modal inputs or outputs, specifying the types of data it

> can process (e.g., text, images, audio).
>
> These descriptions should not be so detailed as to allow replication of the model or to reveal trade secrets that could significantly advantage competitors. The registry should also include provisions for periodic reviews of architectural disclosure requirements to ensure they remain relevant and effective as AI technology evolves.

### Hardware Information

AI labs require huge amounts of specialized hardware for training, testing, deploying, and iterating frontier AI models. Information regarding this physical hardware could be useful for designing and enforcing future AI governance, but also poses particularly difficult tradeoffs for inclusion on a model registry. The hardware in question includes:

- AI Chips, a term which broadly refers to a class of semiconductors that are essential for frontier AI models. These are remarkably specialized for a specific kind of large-scale highly repeatable calculations necessary for the training and execution of neural networks (and thus frontier AI models)[204]. AI chips are typically either more broadly applicable Graphics Processing Units (GPUs) such as Nvidia's H100 chip[205], or more customized Application-Specific Integrated Circuits (ASICs) such as Google's custom TPU chips[206]. Frontier AI models typically require tens to hundreds of thousands of these chips[207], totalling billions of dollars of hardware investment.

- Supporting computational infrastructure such as RAM, high-bandwidth memory, and high-speed networks[208], as well as infrastructure customized for AI such as Nvidia's DGX systems[209].

- Data centers and clusters that contain a large number of AI chips and computational infrastructure[210].

- Supporting physical infrastructure such as cooling systems for the AI clusters[211], power generation or storage[212], security, and so on.

Hardware is becoming a popular focus of AI governance proposals due to its specialization, its role as a bottleneck in producing frontier AI models, and its physical nature, which makes policies easier to implement and enforce[213,214]. Current proposals include chip registration policies[215], tracking compute through chips[216], and many more[217].

Information such as the size and location of critical computing clusters and data centers, detailed lists of hardware used, deployment details (cloud vs. on-site), and so on could improve governments and policymakers' ability to design and enforce effective AI hardware governance in the near future. Some current recommendations for disclosure do recommend sharing such details: see the Institute for AI Policy and Strategy's proposal for coordinated disclosure[218], which recommends collecting information such as the quantity and variety of chips used by AI labs, and the physical location of compute.

However, there are major risks and drawbacks to requiring such detailed hardware information:

- Information on the location of hardware centers could render them targets for physical attacks, sabotage, or espionage. Malicious actors could exploit this information to plan and execute attacks, putting infrastructure and people's safety at risk[219];

- Information on the type and amounts of hardware used could make critical infrastructure more vulnerable to cyberattacks[220], for example by making exfiltration attacks on specific data centers easier to conduct[221];

- Data breaches could lead to the dissemination of hazardous information and increase risks, for example by leaking cutting-edge AI development techniques or enhancing race dynamics[222], each enhancing capabilities.

- The competitive advantage provided by keeping hardware usage strictly private may incentivize labs to resist registry legislation.

There are precedents for governments securely storing sensitive information that could otherwise incentivize or support malicious behavior, such as the US Nuclear Regulatory Commission on nuclear power plants and materials[223] or Chemical Facility Anti-Terrorism Standards, which includes sensitive information on high-risk chemical factories[224]. However, there are also numerous examples of governments' inability to secure sensitive data, such as the 2015 data breach of the US Office of Personnel Management by an advanced persistent threat based in China which affected 22.1 million records[225], or the time a nun and two pacifists bypassed security and accessed a facility holding 100 tons of enriched uranium for several hours[226].

These risks are likely to make AI developers especially resistant to sharing this information. We believe that this domain of information may face significant pushback from labs without providing much immediate benefit to policymakers.

### Our Recommendations

We recommend that model registries should not require detailed information on physical hardware associated with qualifying models due to the increased risk from storing this information, the increased resistance from AI labs if such information was required, and the lack of a clear, immediate advantage for policymakers and governments to access this information.

However, we believe there are some hardware-related questions that could provide useful insights while posing little risk and provoking minimal resistance from AI labs. Specifically:

- AI developers should be required to report the total compute capacity (in FLOP/s) of the hardware clusters used to train and run their models. This provides a proxy for model capability without revealing sensitive details about specific hardware configurations.

- Developers should disclose whether their model is deployed on-premises or via cloud services. If cloud services are used, the provider(s) should be named.

- We also recommend the following, though less strongly than the points above:
  - The registry should require reporting on any significant changes to hardware infrastructure that could significantly increase the total compute capacity.
  - Developers should disclose the total number of AI chips used in training and inference of an AI model and which AI chips they make use of (i.e. manufacturers and models) without revealing proprietary design details.

We recommend that the registry be designed with future adjustment in mind, as it will be increasingly important to adjust to hardware reporting requirements as AI governance frameworks evolve. This should be done in collaboration with AI labs and cybersecurity experts to ensure the information requested is relevant and proportionate to the evolving landscape of AI development and associated risks.

## Model Security

Unrestricted access to frontier models or sensitive information such as their model weights or the results of capability evaluations and risk assessments produces significant risks. Malicious actors can use this information to recreate or misuse a model, or to sabotage or steal intellectual property from competitors, as discussed in _Should the information in the registry be confidential?_. Competing states have strong incentives to exfiltrate model weights and other sensitive information. AI safety advocates and AI labs are aligned in a desire to maintain both physical protections and cybersecurity around frontier models and sensitive information.

A model registry could be used by the government to ensure that AI labs are establishing and maintaining sufficiently rigorous security and cybersecurity measures. There are analogous precedents in other high-risk industries where governments decide that certain technologies are too dangerous to leave security entirely to private discretion. For example, nuclear power plants must provide security plans to the Nuclear Regulatory Commission, including physical protection and cybersecurity programs[227]. Similarly, pharmaceutical companies handling controlled substances must register with the DEA and implement strict physical security and inventory controls[228].

[228] ENISA lists 6 broad categories of asset: Data, model, actors, processes, environment/tools and artifacts. These categories are broken down into more detail in their AI Cybersecurity Challenges report, 23.

For AI models, this could include providing details of the measures taken to secure model weights, architecture, training data and source code, as well as measures taken to prevent the misuse of legitimate APIs and any emergency response plans if any sensitive information is exfiltrated[229].

Mature cybersecurity standards exist in other domains, but AI presents some unique challenges. Across a model's lifecycle, attackers can target a range of different assets, from the model itself to the actors involved in its development and deployment[230]. These assets exist in a complex ecosystem of evolving techniques, deployment scenarios, supply chains and associated fields (such as facial recognition and robotics)[231]. Best practices in AI cybersecurity continually change and rapid responses to zero-day vulnerabilities are necessary[232]. AI itself is accelerating changes in the field of AI cybersecurity by making cyber operations broadly more sophisticated and accessible[233]. Competing entities are strongly motivated to access other entities' models and data to advance their own model capabilities.

Due to the complex landscape, it's difficult to establish a consensus on standards for sufficient cybersecurity in frontier AI labs. Several major standard development organizations have established dedicated bodies for AI cybersecurity, including ETSI, NIST, IEEE, CEN/CELEC and ISO[234]. AI Labs are also implementing responsible scaling policies that include implementing security measures proportional the risks presented by a model[235]. However, these have not been tested in real-world scenarios, and the rapid development of an increasingly complex AI cybersecurity threat landscape is likely to reveal unforeseen vulnerabilities and challenges.

Additionally, given the potential role of AI in CBRN threats, as discussed in *What thresholds should a model exceed to qualify for inclusion? - High-risk domains*, it would be reasonable to subject these systems to similar security requirements as in other domains that pose similar risks. When an AI model is determined to pose a significant CBRN risk, governments should draw on standards used to control access to information that has significant implications for national security. For example, they may conduct background checks and record who has access to the AI model, as is done for people with access to select biological agents in the US[236].

Implementing mandatory cybersecurity measures and governmental oversight would impose significant responsibilities on both AI labs and registry administrators. Maintaining state-of-the-art cybersecurity standards necessitates dedicated teams of full-time cybersecurity professionals. While established frontier AI labs typically have such teams in place, the rapidly

---

[236] A model registry collecting information on existing cybersecurity practices would itself be a key source of information for improving current and future standards.

[241] For a list of these organizations see Study of Research and Guidance on the Cyber Security of AI 16-18

expanding size and capability of AI models may require smaller organizations to meet similar standards in the future, leading to substantial operational and financial burdens. Excessively rigorous cybersecurity requirements might impede innovation, slow research and development, and add considerable overhead to AI development processes.

Similarly, verification of cybersecurity standards would require significant overhead from government agencies. Agencies would need to maintain a team of cybersecurity experts, or contract with external organizations to conduct cybersecurity audits.

### Our Recommendations

We recommend that a model registry should require that model developers describe the cybersecurity measures taken to protect key components of their AI model, including model weights, proprietary training data, and the source code of AI models. In addition, model developers should report measures taken to protect personally identifiable information used in the training of AI models, as is required by legislation such as the GDPR[237].

We also recommend the selection and adoption of a standardized framework for evaluating the cybersecurity of AI models[238,239]. Current examples of standardized frameworks for AI models include RAND's Security Levels for AI model weights[240], and Deepmind's Frontier Safety Framework levels for Security Mitigations[241]. Frameworks should address serious threats across the full lifecycle of the model, considering model design and development (including requirement analysis, data collection, training, testing, integration), installation, deployment, operation, maintenance and disposal[242].

Within such a framework, we recommend that governments should formally identify a set of acceptable standards that are deemed appropriate to measure the cybersecurity of key components of AI models. These standards should be drawn from the work done by standard-setting organizations with respect to AI specifically[243], as well as other mature cybersecurity standards such as ISO/IEC 27001[244], FISMA[245], and NIST[246] and MITRE[247]. These standards could serve as a foundation for developing mandatory cybersecurity requirements in the future.

### Model Evaluations and Risk Assessments

Regulation of technology and commercial products often relies on safety and security evaluations. For example, developers may need to demonstrate to regulators that their products pass standardized assessments, as with the FDA's preclinical and clinical research stages before new drugs are approved[248].

Many experts believe such assessments will be vital to long-term AI safety. AI developers do already conduct assessments to measure risks, limitations, and performance before deployment; these include typical risk assessments, similar in form and scope to other industries, as well as AI-specific evaluations, sometimes called *evals*. However, risk assessments from other industries aren't sufficient to prove that a frontier model is safe, and the science of evals is nascent.

**The science of evaluating frontier AI models**

There are four broad categories of AI model evaluation:

- **Capability** evaluations (or performance evaluations, benchmarks, etc) are the broadest category, and are used to assess how well the model accomplishes particular tasks. These are often standardized and can be used to advertise the capabilities of the model[249].

- **Safety** evaluations assess the potential for AI models to cause unintended harm or lead to harm through misuse[250]. Safety evaluations are common, but the flexibility and range of capabilities of frontier AI make such models extremely difficult to evaluate sensitively. For this reason, many researchers are now developing frontier AI-specific safety evaluations, though current evals are not sufficient to guarantee safety[251].

- **Security** evaluations identify cybersecurity vulnerabilities that would let malicious actors remotely access the model, misuse the model, access the model weights, and so on.

- **Alignment** evaluations assess how well the goals of the model align with the goals of users (and humanity, more broadly). For example, there is some misalignment between LLMs and their users – LLMs are merely trained to predict the next word in a sequence, not to predict truthful sentences, and as a result oftentimes hallucinate false responses.

Frontier AI models are uniquely difficult to robustly and sensitively evaluate. These models are incredibly flexible, easily customizable, and undergo frequent and unpredictable innovation. Two different people with different aims and different skills could use customized versions of GPT-4 to achieve wildly different outcomes – for example, to write an essay and to generate instructions for constructing bioweapons[252].

This poses an issue for governments. While safety testing is often mandatory in other industries, we lack the tools to demonstrate model safety. Governments could instead rely on evaluations that are *specific* rather than *sensitive*, i.e. tests for specific known threats that aren't designed to measure overall safety. These could be useful as an initial safety check, and as a prerequisite for more advanced and sensitive evaluations. However, this does incur a risk of safety-washing[253], in which the public believes a model is safe despite insensitive testing.

Recent regulations have tackled this issue by requiring risk assessments and safety evaluations, with steps taken to mitigate risk, but without specifying any particular standards. For example, Article 55 of the EU AI Act requires providers of general-purpose AI models with systemic risk to perform model evaluations in accordance with the state of the art and to assess and mitigate systemic risks[254]. Similarly, the proposed Californian *Safe and Secure Innovation for Frontier Artificial Intelligence Models Act* would require developers of qualifying models to implement safety and security protocols and to publish a redacted copy of this protocol[255].

Despite the current lack of tools, we can be optimistic that evaluations will improve in the next few years. Independent research organizations such as METR[256] and Apollo[257] are rapidly developing and sharing results[258] on model evaluations, sometimes called evals. For more detail on how these nascent evaluations actually work in practice, see our State of the AI Regulatory Landscape[259] in which we break down one of *Model Evaluation and Threat Research*'s pilot studies.

Governmental bodies such as the UK AI Safety Institute are also focusing on developing better evaluations[260], in collaboration with independent research organizations. There is a strong demand for robust, sensitive evaluations, and indeed many proposals for long-term AI safety discuss applying such evaluations to AI models, such as responsible scaling policies[261]. The AI Bill of Rights calls for pre-deployment testing, risk identification & mitigation, and ongoing safety monitoring[262], and the US Executive Order is enacting measures to develop evaluation techniques and infrastructure[263].

### Sharing the results of in-house evaluations

Some recent AI policies such as the US Executive Order on AI[264] find a middle ground by requiring developers to share the results of in-house evaluations and red-teaming exercises. The UK AI Safety Institute currently requests AI labs to voluntarily share data[265]. These approaches may not stop unsafe models being deployed, but they offer governments more insight into how AI developers measure safety, alignment, and security. They improve accountability and give policymakers a greater opportunity to interpret and develop effective evaluations.

While it may not yet be practical to include standardized evaluations as part of the registration process, we could still require AI labs to release the results of any evaluations they do conduct.

### Our Recommendations

We do not currently recommend that a model registry should require AI developers to conduct any particular or standardized evaluations. We do, however, recommend that developers registering a model should be required to share details of the nature and results of their evaluations, following the approach described in the US Executive Order on AI[266]. Secondly, we recommend that governments should continue to invest in the development of high-quality AI evaluations for safety, security, capability, and alignment. These, as well as traditional risk assessments, should be gradually incorporated as necessary requirements for deploying a model publicly. Once more comprehensive and vetted model evaluations exist, we would recommend that these evaluations be required for inclusion into a model registry

We provide an example of the types of information a model registry could request AI developers share from previously conducted model evaluations:

1. **Evaluation types:** Developers should specify which types of evaluations were conducted (capability, safety, security, and/or alignment evaluations).

2. **Evaluation methodologies:** For each evaluation type, provide a brief description of the methodology used, including any standardized benchmarks or custom evaluation frameworks.

3. **Performance metrics:** Report key performance metrics for each evaluation, including both aggregate scores and more granular breakdowns where available.

4. **Instance-level results:** Where possible, provide access to instance-by-instance evaluation results to allow for more detailed analysis.

5. **Red-teaming results:** Summarize the outcomes of any red-teaming exercises, including successful attempts to bypass safety measures or exploit vulnerabilities.

6. **Safety and security risks:** Outline any potential safety or security risks identified during evaluations, along with proposed mitigation strategies.

7. **Alignment insights:** For alignment evaluations, describe any misalignments identified between the model's behavior and intended goals.

## Functions of the Model

### Primary or intended use

Registries often require product developers to provide the purpose or intended use of registered items. For example, the FDA's registry of medical devices[267], the EPA's pesticide registry[268], the EU Clinical Trials Register[269], and the EU's chemical registry[270] all require developers to provide the purpose or intended use of registered items. Such information provides important context, helping administrators understand how products will be used in the real-world and the likely risks and harms that will come with that use. It supports the development of further governance by giving policymakers more information to work with; and it also supports more targeted governance, for example by lowering restrictions for lower-risk uses. In the case of AI, this could enable future governance to be more targeted for high-risk areas, such as biochemical development or cybersecurity and others described in *What thresholds should a model exceed to qualify for inclusion? - High-risk domains*.

However, defining use cases can be complex for AI. AI models *are* trained to optimize a specific well-defined reward function, but this doesn't always match the practical use of the model. For example, LLMs like GPT-4 are trained to *predict* the next word in a string of words, but practically are used to *generate* long strings of text. This makes LLMs hugely flexible, as the generated words can be an essay, instructions for building a bomb[271], or even, with additional support, a series of research papers[272]. However, simple explanations of basic use such as "This model generates strings of text based on prompts" should generally be easy for developers to provide and still provide value for the registry.

### Potential uses

As described above, AI models and LLMs in particular often have capabilities beyond their intended use case. These models can be fine-tuned or prompted by users to have specific uses that developers didn't deliberately train for. These uses are still valuable for a registry for the same reasons described above, but are more complex to describe. A comprehensive account of all the potential uses of an LLM would be an unreasonable requirement for a registry, if even possible. However, AI developers do conduct many safety assessments, performance benchmarks, and experiments with models which can identify alternative uses, such as those in the paper announcing GPT-3[273]. These could make it easy for developers to provide at least some examples of alternative uses of their models.

### Model documentation

AI developers often produce documentation that describes uses of models and guidance for users, in the form of research papers[274], API references[275], starting guides[276], and so on. These are publicly available and often contain additional information on usage, so including this information in a model registry would not be overly burdensome for developers and would be useful for administrators in understanding more about registered models.

This would also match registries in other industries. The FDA's medical device registration asks manufacturers to provide links to labeling and instructions for use[277], for example.

### Our Recommendations

We recommend that a model registry should require developers to describe, in a few sentences of plain language, what each registered model does and what its primary purposes are. We also recommend that developers be asked to describe any major alternative uses identified during development and to provide links to any publicly available documentation on the usage of the model.

## Post-Deployment Monitoring

Post-deployment monitoring is the practice of tracking and monitoring the performance, security, and reliability of a system or application after it has been deployed. This is important in many fields to confirm that the impacts of a product are as predicted and to generate information on making further improvements to the product or how it is deployed.

Monitoring after deployment is especially important for AI systems for both governments and the private sector, since AI systems involve an unusual amount of uncertainty due to the often inscrutable nature of their internal decision-making processes. How AI is used in the wild can reveal emergent behaviors and uses, as well as unexpected interactions with other systems that are difficult to predict in labs[278]. AI labs already monitor a range of metrics on model behavior and use. OpenAI, Anthropic and others use classifier models[279] and abuse pattern capture[280] to identify misuse. Real-time monitoring and input/output logging are both leveraged to spot risks quickly[281], and users are routinely given easily accessible ways to report when a model functions improperly[282]. Some labs have also committed to disclosing vulnerabilities or incidents with other labs[283].

There are also some early efforts in the public sector to monitor live AI models, such as existing model registries in the US, UK, and EU (introduced in _What AI model registries currently exist?_), or the foundation model transparency index[284], which measures models' distribution and impacts. These efforts are less mature than those in the private sector, and these public and private efforts lack coordination. Even between developers and deployers, limited information sharing increases opacity into the systems' performance in the wild for both private and public sectors[285].

In other domains, post-deployment monitoring for governance is conducted directly by governments, such as NHTSA monitoring vehicle safety[286], or regular inspections of nuclear facilities by the US RNC[287]. Alternatively, or additionally, governments can ensure that private entities will take on the monitoring and reporting themselves. For example, the US RNC also requires responsible entities to have arrangements to identify, record, and investigate abnormal incidents, and to notify the Office for Nuclear Registration[288]. More specifically in the context of AI regulation, the EU AI Act requires dutyholders to report any incidents defined as 'serious incidents'[289]. In these latter cases, governments can capitalize on existing monitoring infrastructure and capabilities in industry, but have less oversight in the process. This could risk enabling labs to doctor or withhold information, and also may contribute to a more fragmented regulatory landscape, with a range of different, and possibly disparate reporting mechanisms and standards in use within one jurisdiction.

While nations are making meaningful progress, there are no robust national standards for post-deployment monitoring. For example, the EU AI act requires that developers of high risk models submit a post-market monitoring plan, however the details of what this plan would look like are not projected to be developed until February 2026[290]. Thus, direct post-deployment monitoring of AI systems by governments may be difficult to implement at this stage. Meanwhile, some private labs are taking steps to prepare to inform public authorities in the event of a serious incident[291].

A registry could provide a more mature mechanism for post-deployment monitoring, and support coordination between the public and private sectors. As we have argued previously, a registry should place a minimal regulatory burden on AI developers, and should aim to meet its specific governance objectives. Therefore, while there are hundreds of KPIs that could be monitored, the registry should aim to only measure information that can contribute directly to stated governance objectives (see *The Case for a Model Registry*), such as increasing visibility into risks associated with a system, informing new regulation (e.g RSPs, incident reporting, and licencing based on model capabilities), and supporting enforcement of existing regulation.

### Our Recommendations

**Initially, a model registry should not involve governmental monitoring of AI systems directly after deployment,** but should require labs to share information about their own post-deployment monitoring practices, including:

1. What safety KPIs are being monitored.

[295] You can read more about AI safety and incident reporting in our regulatory review on the topic.

2. Thresholds for when safety KPIs trigger a response.
3. Protocols detailing what responses are triggered by any given incident, including when an incident or vulnerability should be reported to the relevant government bodies.
4. Policies for reviewing post-deployment monitoring practices[292,293].

**A registry should support efforts to develop national standards on post-deployment frontier model monitoring** so that governments can eventually monitor the AI ecosystem more directly, analogous to monitoring practices of the FDA[294], NHTSA[295] (vehicle safety), US environmental Protection Agency[296], and other mature government agencies that use post-deployment monitoring to assess the effectiveness of regulation.

**While registries verify that incidents will be reported to the relevant bodies, registries themselves may not be databases for reporting incidents.** However, the above proposal may support the creation of such a database by further developing norms and unifying standards around incident reporting[297].

# Part III – Implementation of a Model Registry

**Should compliance be ensured by requiring third parties to use only registered models?**

To ensure that AI developers register qualifying models, a practical system to incentivize compliance must be implemented. This can involve financial penalties for companies that fail to comply with registry requirements, which we'll discuss in *Should non-compliant AI developers face financial penalties?* Another powerful mechanism to ensure compliance is injunctive action; eliminating the market for unregistered models by requiring third parties to ensure the models they use are registered.

Such a mechanism has many precedents in analogous cases. For example:

- Know-Your-Customer (KYC) standards in banking[298] protect financial institutions against fraud, money laundering, and terrorist financing by requiring banks to verify customer identities, assess the nature of their activities, and evaluate their funding sources as legitimate and the risk of money laundering as low.

- Professional licensing requirements penalize people who purchase products or services from unlicensed professionals. For example, the US DEA requires healthcare providers and pharmacies to verify that prescriptions are from licensed practitioners before dispensing controlled substances, and face penalties if they fail to do so[299].

- Worker registration systems such as the US E-Verify system[300] require employers to verify the eligibility of their employees to work in the United States. Employers face penalties for knowingly hiring or continuing to employ unauthorized workers.

A requirement for third parties to ensure the models they use are registered would bring AI regulation to the same level as the examples given above. This would decentralize enforcement and spread responsibility across the market ecosystem, creating a strong incentive for timely and accurate registration by AI developers. Unlike direct financial penalties, this mechanism doesn't require a heavy public sector enforcement mechanism; as we've seen recently with the IRS, the impact of public sector enforcement is limited by investment and the number of auditors[301].

Implementing this system would require a publicly accessible component of the registry. This would be similar to the FDA's searchable database of approved drugs[302] and unique Premarket Approval Numbers as identifiers for medical devices[303], while still allowing the confidentiality and security for most information as we advocate for in *Should the information in the registry be confidential?*. The registration process and ability to verify that models are registered should be designed to be as lightweight and efficient as possible.

Alternatively, the registry could provide AI developers with a verifiable "stamp of approval" or unique identifier to embed in their products' user interfaces or include in advertising materials. This would allow users to easily verify a model's registration status without needing direct access to the full registry, analogous to digital certificates for website security, in which SSL/TLS certificates are issued to websites by trusted Certificate Authorities. These certificates provide a padlock icon in web browsers so users can easily verify a site's security credentials[304]. Establishing verification systems like this and ensuring they aren't counterfeitable will be important for long-term governance and this requirement on third parties will spur innovation in that direction.

These requirements could be enforced by fines on individuals or organizations and companies using AI models in their business. These could be graduated, with fixed or relatively small-percentage fines for individuals or minor violations from organizations, up to fines of a substantial percentage of annual turnover for large organizations or major violations. The regulatory body should also have the flexibility to adjust these fines as the AI landscape evolves and the impact and scope of the registry becomes clearer.

Note that requirements on third parties to use only verified models should only apply to models that would qualify for inclusion on the registry. If a third party uses a model that doesn't meet the criteria for inclusion, discussed in *What should qualify for inclusion on the registry?*, they are not obliged to verify that the model is registered.

This enforcement mechanism is likely to be most effective in the business-to-business market. Prosecuting every member of the public who uses a LLM is unlikely to be a good use of public sector resources. However, if developers of phones, search engines, software, and so on face consequences for using or selling access to unregistered models, they have a strong incentive to ensure that the AI developers guarantee their products are compliant. AI developers will be encouraged to guarantee to their large clients that their models are properly registered (or that they demonstrably do not meet the requirements for registration). Such assurances come under representations and warranties[305] common in many industries and contracts.

### Our Recommendations

We recommend that an AI registry should be enforced by requiring third party users of AI models to verify that the models they are using

> have been registered when said model meets the criteria for inclusion on the model registry.
>
> To support this, registry administrators should maintain a publicly available and easily searchable database of registered models, assigning each a unique identifier. Administrators should also issue registered model developers with digital certificates demonstrating compliance and establish a clear visual symbol, stamp, or logo that developers can use to show the public that the model is registered.
>
> This will also require a program of public education or notification to ensure that such third parties are fully aware of the legal requirements on them and that they recognize symbols of compliance and understand how to verify that the models they're using are registered. To allow for such education, these public-facing requirements could be staggered or come into effect 6-12 months after the registry is established.
>
> Individuals or organizations who fail to comply with these requirements should face fines proportional to the scale of use and to annual turnover for large organizations.

### Should non-compliant AI developers face financial penalties?

As introduced in *Should compliance be ensured by requiring third parties to use only registered models?*, the primary enforcement mechanism we recommend is injunctive action, but another way to ensure developers properly register their models is to fine those who don't.

Fines for failure to comply with governance are a common penalty[306,307]. These can generally be either fixed fines or proportional to annual turnover. Fixed fines can either be one-off or, more commonly, periodic until the company complies with regulation. For example, the UK Companies House register of overseas entities fines offending parties through a fixed penalty and/or a repeating daily penalty until they're compliant[308].

However, fixed fines will be difficult to calibrate for AI developers. Some models have been deployed by small teams with limited financial resources for academic purposes, while many of today's largest models are developed by large AI labs with immense financial resources[309]. Instead, penalties could be calculated as a percentage of the company's annual turnover. This approach ensures the impact is proportional to the size and financial capacity of the organization, maintaining a meaningful deterrent effect across the industry. A penalty proportional to annual turnover aligns with regulatory approaches in other high-stakes industries, providing a familiar enforcement framework. This approach would be similar to:

- **Drug registration:** The FDA requires registration of new drugs, and fines

companies when they provide false information or promote drugs for uses not included in registration. In 2009, Pfizer was found to promote off-label use of drugs and were fined $2.3 billion[310] (representing 4.6% of annual turnover[311]).

- **GDPR violations:** The EU's General Data Protection Regulation allows for fines of up to 4% of annual global turnover for the most serious infringements[312]. For example, Meta Ireland was fined €1.2 billion in 2023 for violating data transfer requirements (representing around 2% of annual turnover)[313].

This regulatory approach typically incorporates escalating fines for repeated or particularly severe violations. For example:

- **Environmental Protection Agency (EPA) emissions reporting:** Companies must report certain emissions data to the EPA, with harsh fines for egregious violations. In 2015, Volkswagen was fined $4.3 billion (equivalent to around 1.9% of their sales revenue that year[314]) for violating the Clean Air Act by installing software to cheat emissions tests, with penalties based on the number of vehicles affected[315].

### Our Recommendations

We recommend implementing a system of fining AI developers some percentage of annual turnover or a daily fixed fine for non-compliance with registry requirements, with variation depending on the severity and frequency of violations. However, the primary incentive should be enforced through preventing sales of unregistered models, as discussed in _Should compliance be ensured by requiring third parties to use only registered models?_.

## To what degree should the administration of a model registry be out-sourced to third parties?

Governments establishing registries will need to choose whether to develop them and their surrounding infrastructure within government agencies or whether to outsource some or all of this work to third parties.

Such outsourcing is not uncommon for government regulation of complex private goods or technologies. For example, the US government works with third parties like Moody's and S&P for analyzing the value and risk of bonds, securities, and other financial instruments[316]. Similarly, the FAA delegates some aspects of safety monitoring to some airlines, allowing them to self-certify specific aspects of aircraft design and production[317]. Outsourcing is appealing in these cases, as these independent organizations can complete work more quickly and efficiently using existing industry expertise and economies of scale, thereby reducing government costs without losing the value of the work.

However, outsourcing has its costs; it can lead to conflicts of interest, a lack of accountability, and insufficient oversight or even corruption. Indeed, the examples above have both lead to controversy, as reliance on credit ratings contributed to the 2008 subprime mortgage crisis[318]. Similarly, Boeing has recently had a string of technical failures and crashes for which investigators found shortcomings in Boeing's certification with the FAA to be responsible[319].

We expect AI governance to expand massively in scope in the coming years, and that will require a growing infrastructure and technical expertise within governments to design and enforce regulation. Outsourcing these early regulatory mechanisms will delay and reduce the government's capacity for technical AI work; developing registries in-house could instead be an excellent opportunity for early development of infrastructure and acquisition and training of people with technical expertise in AI. Governments should maintain ownership over and visibility into this work to strengthen policy-making and expertise[320].

There should be a tight loop between those monitoring AI and those writing AI policy. This will improve the government's technical expertise, strengthen the government's relationship with stakeholders in academia and industry, and lead to the development of infrastructure that can be reused for different purposes in the future. A comprehensive AI registry, developed and administered by the government, could be a foundational tool for further AI regulation.

### Our Recommendations

In light of these tradeoffs, we recommend that the registry design proposed should be established and maintained within a governmental agency, with minimal outsourcing to third-parties in order to build the government's technical expertise and capacity to manage AI governance.

### At what administrative level should a model registry be implemented?

A model registry could be established and enforced within governments at different scales, including:

- At the city or state-level, as with New York[321], Amsterdam[322], and Helsinki's[323] algorithm registries used by each cities' local government;
- Nationally, as with China's model registry[324] and the FDA's registries of

food products and medical devices[325];

- Internationally, such as the contact database for high-risk AI systems proposed in article 49 of the EU AI Act[326], or the REACH register of chemical suppliers in the EU[327].

Registries can also be maintained internally by individual organizations – for example, by international organizations such as the International Atomic Energy Agency[328] (maintaining the Nuclear Material Accounting Database), or disease-specific patient registries maintained by non-profit organizations or academic medical centers[329].

Each approach has advantages and drawbacks. International organizations have wider scope and greater political support, but also necessarily involve negotiation between many more stakeholders. Small, local registries can be more efficient and tailored to local laws and regulations, with fewer stakeholders. This makes them more likely to be established early. However, they may have a narrow scope and limited capacity to enforce compliance.

The registry we propose in this report could be effective at many scales, but it does require enforcement. In particular, we propose in *Should compliance be ensured by requiring third parties to use only registered models?* and *Should non-compliant AI developers face financial penalties?* that third parties should be required to ensure that the AI models they use are registered and that non-compliant developers could also be fined. This requires major regulatory influence and judiciary power over a range of actors and stakeholders that would likely be difficult below a national administration.

Note also that a national registry does not limit international collaboration on the issue. Sharing information, enforcement, and responsibilities across nations and larger governmental groups such as the EU and UN would be a powerful way to improve global capacity building, and ensure effective long-term governance. AI models and their impacts transcend national borders, and AI risks such as misuse by malicious actors or the development of dangerous weapons are global in scope and will require a coordinated international response. Even if registries are implemented nationally, governments should work towards mechanisms for international information sharing, coordination, and joint decision-making. Many governmental research bodies are already collaborating across borders, such as the collaborations between the UK[330], US[331], and Canadian[332] AI Safety Institutes and the subsequent commitment between ten countries and the EU at the AI Seoul Summit[333] to build an international network of collaborative AI research institutes.

### Our Recommendations

We recommend that the model registry proposed in this report should be administered at a national level to ensure it can be effectively enforced.

Additionally, we recommend that governments should develop long-term plans to coordinate international standards used in the creation of model registries to promote interoperability, and foster international collaboration by leveraging information collected by model registries.

## Should the information in the registry be confidential?

Registries vary in how much information is publicly available. Some registries are fully accessible to the public, while others only allow certain individuals such as governmental officials to access some or all of the information.

Public registries provide the most transparency and accountability. For example, public registries allow US consumers to verify that the medicine they buy is government-approved and compliant with safety standards and government, to find equivalent alternatives to expensive drugs[334], and to learn whether particular food products may cause adverse health consequences[335].

Registries use confidentiality and information security for several reasons, but most often because the information is too sensitive for the public or for competitors to access. For example, the FDA's Drug Approval Database[336] provides public access to basic drug information and approval statuses, but maintains the confidentiality of proprietary details, such as clinical trial data, to protect intellectual property. Similarly, the EU REACH[337] requires companies to register chemicals they manufacture or import. Basic information, like a chemical's classification and labeling, is made public to promote safe use, but the full composition and manufacturing processes are kept confidential to protect intellectual property.

Information can also be hazardous to share with the public, as malicious actors could use information to cause harm. For example, the The International Atomic Energy Agency maintains the Nuclear Material Accounting Database[338], tracking quantities and locations of nuclear material worldwide to ensure peaceful use and prevent weaponization. Most information is accessible only to authorized IAEA staff and member states, as public knowledge could facilitate theft or sabotage, enable illicit acquisition of weapons capabilities, or reveal proprietary details about power facilities. Similarly, the Wassenaar Arrangement[339] on export controls for weapons and dual-use technology maintains confidentiality of specific export control licenses and transactions to prevent the proliferation of sensitive technologies. General policy information and member commitments are publicly available, but detailed data on specific transactions is kept private to protect national interests and international security.

An AI model registry is likely to contain both commercially sensitive[340] and potentially hazardous information[341]. Information that could be hazardous to share publicly could include:

- The location of hardware, such as data and compute centers, and the identities of people with access to the model and training data (which could be targets for sabotage, harm, or theft).

- The details of a model's structure that could lead to race dynamics or represent commercial advantages, such as algorithm design, model size, how much compute and training data was used to train it, the details of its training algorithms, and so on.

- The results of capability tests, risk assessments, and surveys of alternative uses could give malicious actors insight into how to achieve specific hazardous capabilities, misuse or jailbreak a model, or exploit security vulnerabilities.

- Sharing data could also encourage race dynamics in which developers rush to be the first to reach certain milestones, potentially by disregarding or weakening safety features[342].

Furthermore, an entirely (or near-entirely) private AI registry would be easier to design and enact quickly and efficiently, for several reasons:

- Private registries face less public scrutiny.

- Sharing information publicly is more complex legally, as some information may be protected by existing laws and require upfront legal investigation, negotiation, and resolution.

- Private registries would face less pushback from AI labs providing this information.

### Our Recommendations

We propose that model registries should initially be maintained confidentially and securely, with access only granted to approved individuals within the government. The sole exception is that the public should be able to easily verify whether a particular developer has registered a particular model, to let consumers easily verify the models they're using are compliant. To facilitate this, the model registry should expose a portal that allows consumers to search for a

> model and determine if it has been registered.
>
> In addition to duties of confidentiality, the registry needs to be secure to physical and cybersecurity attacks. As described above, the information in the registry is valuable to malicious actors. Governments may wish draw inspiration from international standards such as ISO/IEC 27001[343] for information security. As a reference, we discuss cybersecurity standards more broadly in *Model Security*.

# Conclusion

AI has advanced dramatically in the last decade, and its impact on our everyday lives, our economy, and our society is likely to continue growing. This rapid development has outpaced governmental capacity to establish basic insight and design effective regulation for AI, in line with insight and regulation in other industries.

Experts disagree about the future of AI. However, few, if any, expect AI to be less prominent in a decade than it is today, and its prominence today already warrants basic governmental oversight to ensure public safety and economic stability. Registries are a standard governmental tool to establish such oversight and to inform future policy-making.

We recognize the need for lightweight and efficient governmental oversight, and so our proposal minimizes the burden on both developers and governments by recommending injunctive action in the market as the primary mechanism to ensure compliance. We recognize the value of innovation and the need for care when dealing with commercially sensitive information. We recognize the need for confidentiality and careful protection of hazardous information. We recognize the difficulty developers face in evaluating the capabilities and risks of their models.

Crucially, though, we recognize that AI development will have a huge impact on society in the coming decades. Governments need to establish basic insight, and our proposal grants that insight without undue burden or risk.

We urge policymakers and AI developers to collaborate in implementing national model registries, as they offer a critical first step towards responsible AI governance that balances innovation with public safety.